\documentclass[11pt]{article}
\usepackage{amsmath}
\usepackage{amssymb}
\usepackage[dvips]{graphicx}
\begin{document}
\title{New Models of General Relativistic Static Thick Disks}
\author{D. Vogt\thanks{e-mail: danielvt@ifi.unicamp.br}\\
Instituto de F\'{\i}sica Gleb Wataghin, Universidade Estadual de Campinas\\
13083-970 Campinas, S.\ P., Brazil
\and
P. S. Letelier\thanks{e-mail: letelier@ime.unicamp.br}\\
Departamento de Matem\'{a}tica Aplicada-IMECC, Universidade Estadual\\
de Campinas 13083-970 Campinas, S.\ P., Brazil}
\maketitle
\begin{abstract}
New families of exact general relativistic thick disks are constructed
using the ``displace, cut, fill and reflect'' method. A class of functions used
to ``fill'' the disks is derived imposing conditions on the first and second 
derivatives to generate physically acceptable disks. The analysis
of the function's curvature further restrict the ranges of the free
parameters that allow phisically acceptable disks. Then this class
of functions together with the Schwarzschild metric is employed
to construct thick disks in isotropic, Weyl and Schwarzschild 
canonical coordinates.
In these last coordinates an additional function must be added to one of the
metric coefficients to generate exact disks. Disks in isotropic and 
Weyl coordinates satisfy all energy conditions, but those in Schwarzschild
canonical coordinates do not satisfy the dominant energy condition.

\end{abstract}
\section{Introduction}

Exact solutions of Einstein's field equations with axial symmetry play an 
important role in the astrophysical applications of general relativity, since 
the natural shape of an isolated self-gravitating fluid is axially symmetric. 
In particular, disk-like configurations of matter are of great interest, since they 
can be used as models of galaxies and accretion disks. 

Solutions for static thin disks without radial pressure were first studied by 
Bonnor and Sackfield \cite{Bonnor1}, and Morgan and Morgan \cite{Morgan1}, and with 
radial pressure by Morgan and Morgan \cite{Morgan2}. Several classes of exact solutions 
of the Einstein field equations corresponding to static thin disks with or without radial 
pressure have been obtained by different authors \cite{Lynden-Bell}-\cite{Garcia}.
Thin rotating disks that can be considered as a source of the Kerr metric were presented 
in \cite{Bicak3}, while rotating disks with heat flow were studied in \cite{Gonzalez1}. 
Also thin disks with radial tension \cite{Gonzalez2}, magnetic fields \cite{Letelier2} 
and magnetic and electric fields \cite{Katz1} were considered. The nonlinear superposition 
of a disk and a black hole was first obtained by Lemos and Letelier \cite{Lemos2}. Perfect
fluid disks with halos \cite{Vogt1} and charged perfect fluid disks \cite{Vogt2} were 
also studied. For a survey on self gravitating relativistic thin disks, see for instance \cite{Karas}.

In the works cited above an inverse style method was used to solve the Einstein equations, 
i.\ e., the energy-momentum tensor is computed from the metric representing the disk. Another 
approach to generate disks is by solving the Einstein equations given a source (energy-momentum 
tensor). This has been used by the Jena group to generate several exact solutions of 
disks \cite{Klein1}-\cite{Klein6}.

Even though in a first approximation thin disks can be used as usefull models of galaxies, 
in a more realistic model the thickness of the disk should be considered. The addition of a 
new dimension may change the dynamical properties of the disk source, e.\ g., its stability. 
Thick static relativistic disks in various coordinate systems were presented in \cite{Gonzalez3}.
The method used to construct the thick disks is a generalization of the well known 
``displace, cut and reflect'' method used to generate thin disks from vacuum solutions 
of Einstein equations. This generalization adds a new step and thus can be named 
``displace, cut, \textit{fill} and reflect'' method. In \cite{Gonzalez3} a particular
function with properties that will be discussed later was used to ``fill'' the disks. 

In this article we present a class of the functions mentioned above and use them 
together with the Schwarzschild metric to construct more models of exact relativistic
thick disks. In Sec. \ref{sec_newt} we discuss the main idea of the  ``displace, cut, fill and reflect'' method
and use the Newtonian Kuzmin-Toomre disk to put constraints on the parameter in the class
of ``fill'' functions so that the disks are physically acceptable. Then in Sec. \ref{sec_iso} we
take the Schwarzschild solution in isotropic cylindrical coordinates to generate thick disks. These 
disks have equal radial and azimuthal pressures but are different from vertical pressures. 
In Sec.\ \ref{sec_weyl} the same procedure is repeated in Weyl coordinates. The resulting 
disks have radial tensions that have the same modulus as the vertical pressures, azimuthal 
tensions near the center and azimuthal pressures for larger disk radii. In Sec.\ \ref{sec_schw} 
thick disks are constructed in canonical Schwarzschild coordinates, which were not 
previously studied. Here an additional 
function must be added in order to generate exact disks. The disks show similar 
characteristics of those constructed in Weyl coordinates, but radial tensions are 
different from vertical pressures. Finally, in Sec.\ \ref{sec_discuss}
we summarize the main results.  

\section{Newtonian Thick Disks} \label{sec_newt}

\begin{figure}
\centering
\includegraphics[scale=0.7]{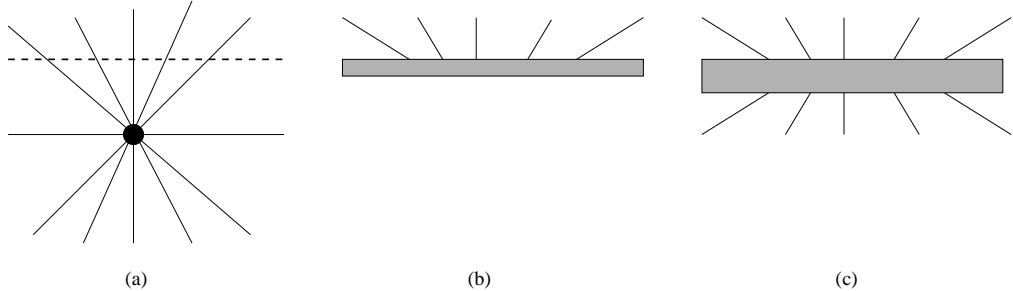}
\caption{Construction of a thick disk by the ``displace, cut, fill and reflect'' method. In (a) the space with a singularity 
is displaced and cut by a plane (dashed line). In (b), after disregard the part with singularities, a thick shell is put below 
the plane. In (c), the resultant configuration is reflected on the bottom surface of the shell.} \label{fig_1}
\end{figure}

In order to construct a thick disk we use a modification of the ``displace, cut and reflect'' 
method, due to Kuzmin \cite{Kuzmin}. It can be divided in the following steps (Fig.\ \ref{fig_1}): first, choose a 
surface (usually the $z=0$ plane) that divides the usual space in two parts: one with no singularities or sources, and
the other with singularities. Second, disregard the part of the space with singularities. Third, put a thick 
shell below the surface. Fourth, use the bottom surface of the shell to make an inversion of the 
nonsingular part of the space. The result will be a disk of infinite extension and thickness twice of 
the shell. In the original method the surface that divides the space is used to make an inversion, resulting 
in a space with a singularity that is a delta distribution with support on $z=0$. Because of the additional
step, the modification can be named ``displace, cut, fill and reflect'' method.

Mathematically the procedure is equivalent to make the transformation $z \rightarrow h(z)+b$, where 
$b$ is a constant and $h(z)$ an even function of $z$. In order to generate thick disks with well defined 
properties, the function $h(z)$ and its first derivative should be continuous in the region $-a \leq z \leq a$,
where $2a$ is the disk thickness. Furthermore, the first and second derivatives of $h(z)$ with respect to 
$z$ should be chosen such that the mass density of the disk (a) is non-negative everywhere and (b) is a 
monotonously decreasing function of $r$ and $z$. In the case of Newtonian gravity, the potential $\Phi(r,z)$ satisfies
the Laplace equation
\begin{equation} \label{eq_laplace}
\Phi_{,rr}+\frac{\Phi_r}{r}+\Phi_{,zz}=0 \mbox{.}
\end{equation}
After we make the transformation $z \rightarrow h(z)+b$, Eq.\ (\ref{eq_laplace}) leads to
\begin{equation} \label{eq_laplace_trans}
\nabla^2\Phi= h^{\prime \prime}\Phi_{,h}+(h^{\prime 2}-1)\Phi_{,hh} \mbox{,}
\end{equation}
where primes indicate differentiation with respect to $z$. From Eq.\ (\ref{eq_laplace_trans}) we
have that, if $|h^{\prime}|=1$ and thus $h^{\prime \prime}=0$ when $|z| \geq a$, the mass density vanishes outside
the disk. For $|z| \leq a$, we get from Poisson equation
\begin{equation} \label{eq_rho_n1}
\rho=\frac{1}{4\pi G}\left[ h^{\prime \prime}\Phi_{,h}+(h^{\prime 2}-1)\Phi_{,hh} \right] \mbox{.}
\end{equation}

If we start from an even and positive polynomial for $h^{\prime \prime}(z)$, we arrive at a class of functions
$h(z)$ that satisfy all the requirements stated above given by
\begin{equation} \label{eq_h} 
h(z)= \begin{cases}
-z+C, & z \leq a, \\
Az^2+Bz^{2n+2}, & -a \leq z \leq a, \\
z+C, & z \geq a,
\end{cases}
\end{equation}
with
\begin{equation*}
A =\frac{2n+1-ac}{4na} \text{,} \quad B=\frac{ac-1}{4n(n+1)a^{2n+1}}, \quad
C =-\frac{a(2n+1+ac)}{4(n+1)} \mbox{.}
\end{equation*}
Here $n=1,2,\ldots$, and $c$ is the jump of the second derivative on $z=\pm a$. 
The special case when $ac=1$ was considered in \cite{Gonzalez3}.
As an example we consider the gravitational potential of a mass point in cylindrical 
coordinates
\begin{equation}
\Phi= - \frac{Gm}{\sqrt{r^2+z^2}} \mbox{.}
\end{equation}
By doing the transformation $z \rightarrow h(z)+b$ in the previous potential and 
using Eq.\ (\ref{eq_rho_n1}), we obtain the mass density
\begin{equation} \label{eq_rho_n2}
\tilde{\rho}=\frac{\tilde{m}}{4\pi}\left[ \frac{\tilde{h}^{\prime \prime}(\tilde{h}+\tilde{b})+
\tilde{h}^{\prime 2}-1}{R^3}+\frac{3(1-\tilde{h}^{\prime 2})(\tilde{h}+\tilde{b})^2}{R^5} \right] \mbox{,}
\end{equation}
with $R^2=\tilde{r}^2+(\tilde{h}+\tilde{b})^2$ and the variables and parameters 
were rescaled in terms of the disk half-thickness: $\tilde{r}=r/a$, $\tilde{h}=h/a$,
$\tilde{b}=b/a$, $\tilde{m}=m/a$, $\tilde{\rho}=a^2\rho$ and with $\tilde{c}=c/a$ and 
$\tilde{z}=z/a$ in Eq.\ (\ref{eq_h}). The mass density will 
be positive if 
\begin{equation} \label{eq_cond_n}
\tilde{h}^{\prime \prime}(\tilde{h}+\tilde{b})+\tilde{h}^{\prime 2}-1 \geq 0 \mbox{.}
\end{equation}
When $\tilde{z}=0$, condition (\ref{eq_cond_n}) reduces to
\begin{equation}
\tilde{b}\frac{2n+1-\tilde{c}}{2n}-1 \geq 0 \rightarrow \tilde{b} \geq \frac{2n}{2n+1-\tilde{c}} \mbox{,}
\end{equation}
and $\tilde{c}<2n+1$ to make $\tilde{b}$ positive. When $\tilde{z}=0$, (\ref{eq_cond_n}) 
gives $\tilde{c}(\tilde{h}+\tilde{b})\geq 0$, or $\tilde{c} \geq 0$. Thus the parameter $\tilde{c}$ is
restricted to $0 \leq \tilde{c} < 2n+1$.

The total mass $\tilde{\mathcal{M}}$ of the disk is easily calculated
\begin{equation}
\tilde{\mathcal{M}}= \int_0^{2\pi} \int_{\tilde{z}=-1}^{1} \int_{\tilde{r}=0}^{\infty}
\tilde{\rho} \, \tilde{r} \, \mathrm{d}\tilde{r} \, \mathrm{d}\tilde{z} \, \mathrm{d}\varphi =\tilde{m} \mbox{,}
\end{equation}
thus the disks have finite mass.
\begin{figure}
\centering
\includegraphics[scale=0.55]{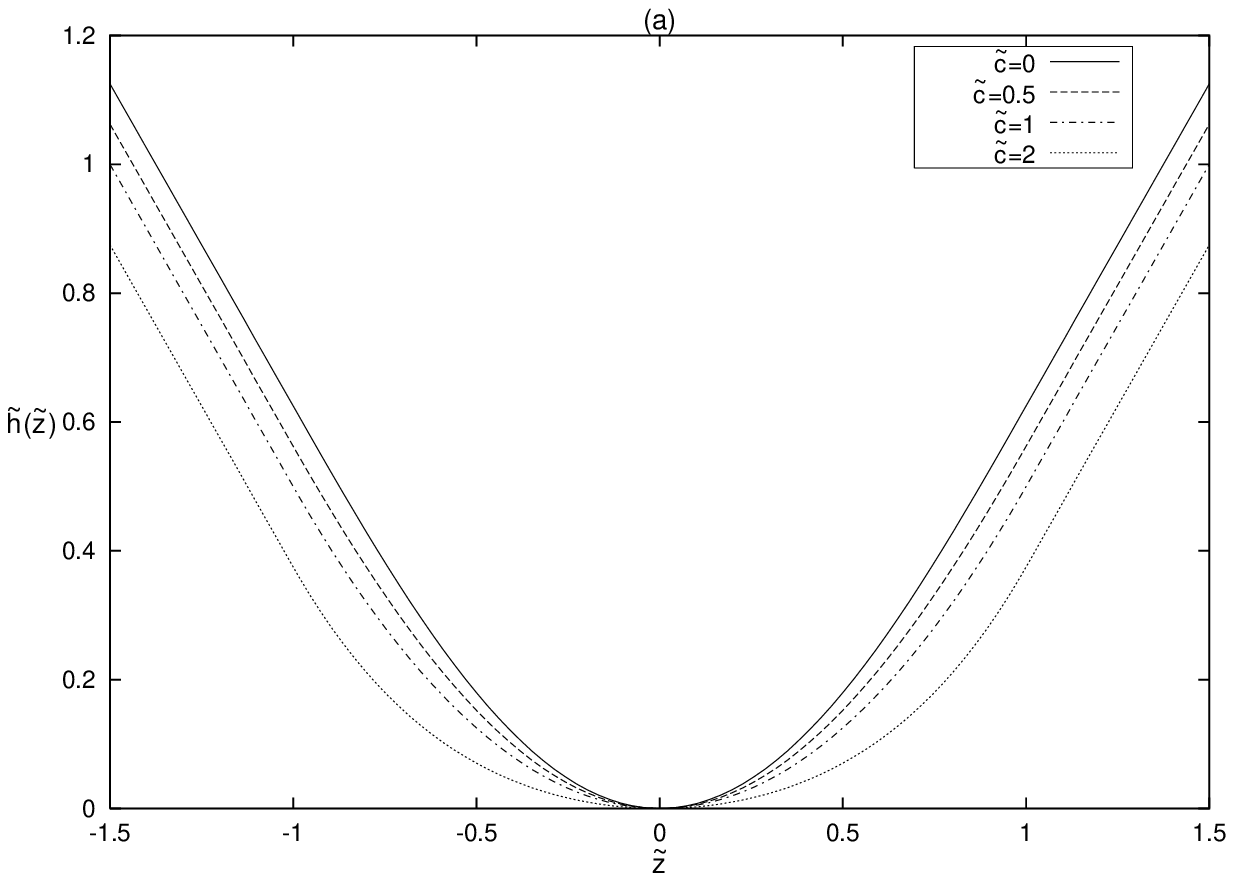}%
\includegraphics[scale=0.55]{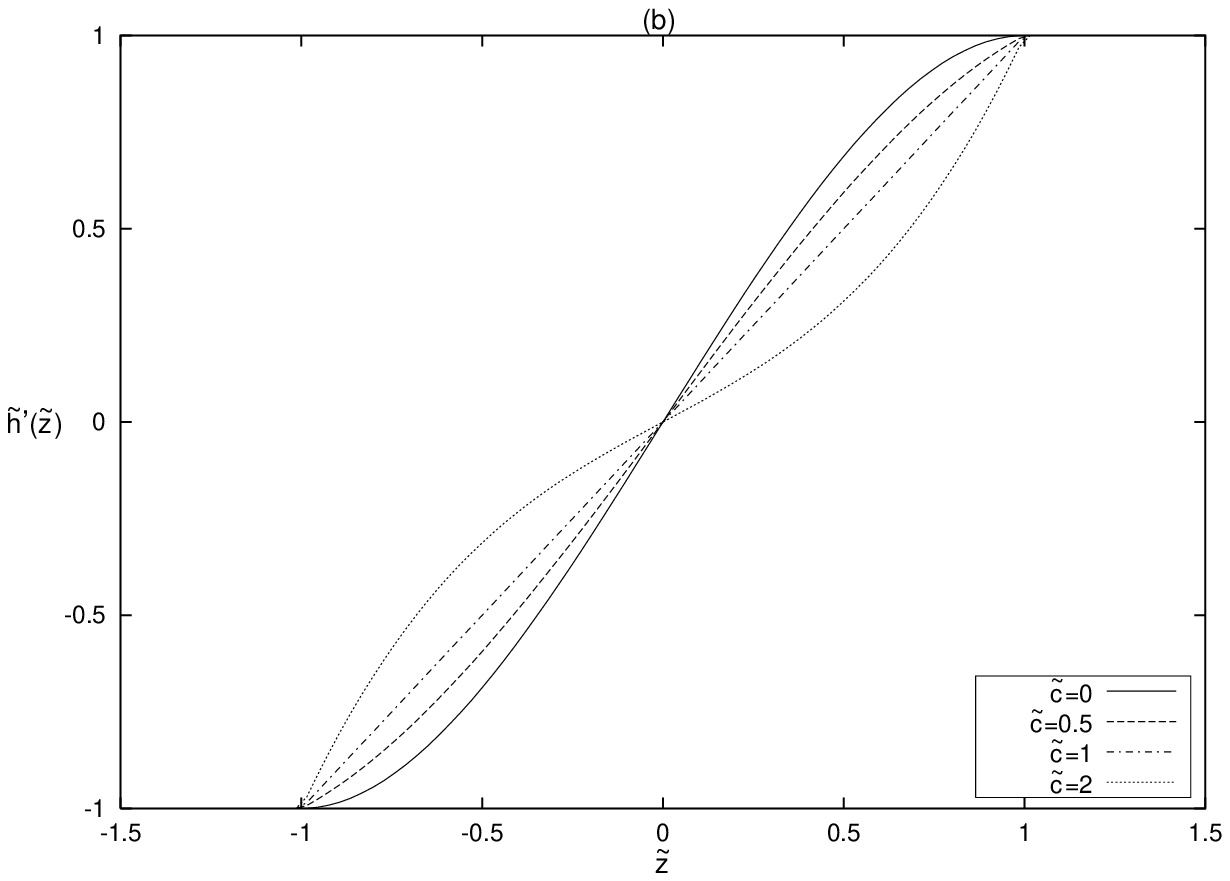}\\
\includegraphics[scale=0.55]{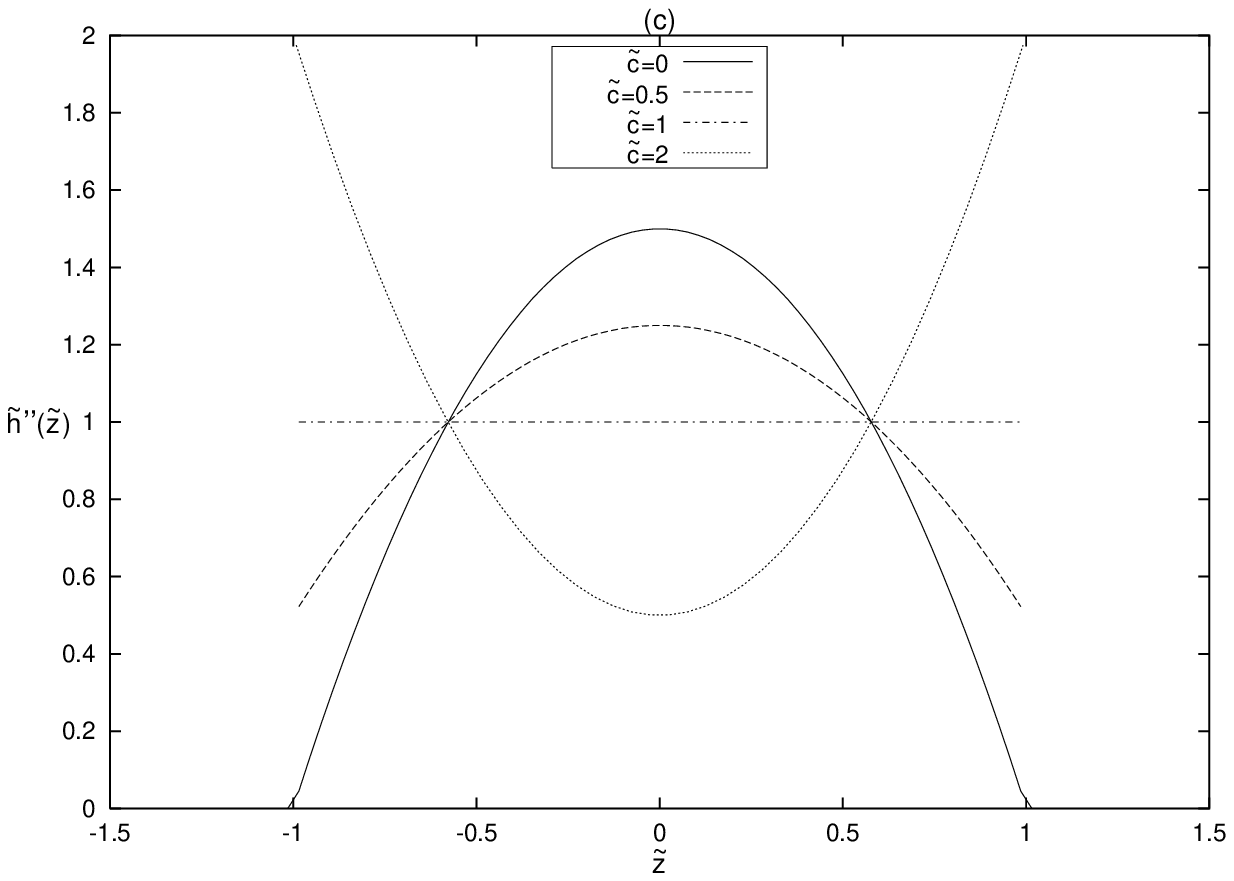}%
\includegraphics[scale=0.55]{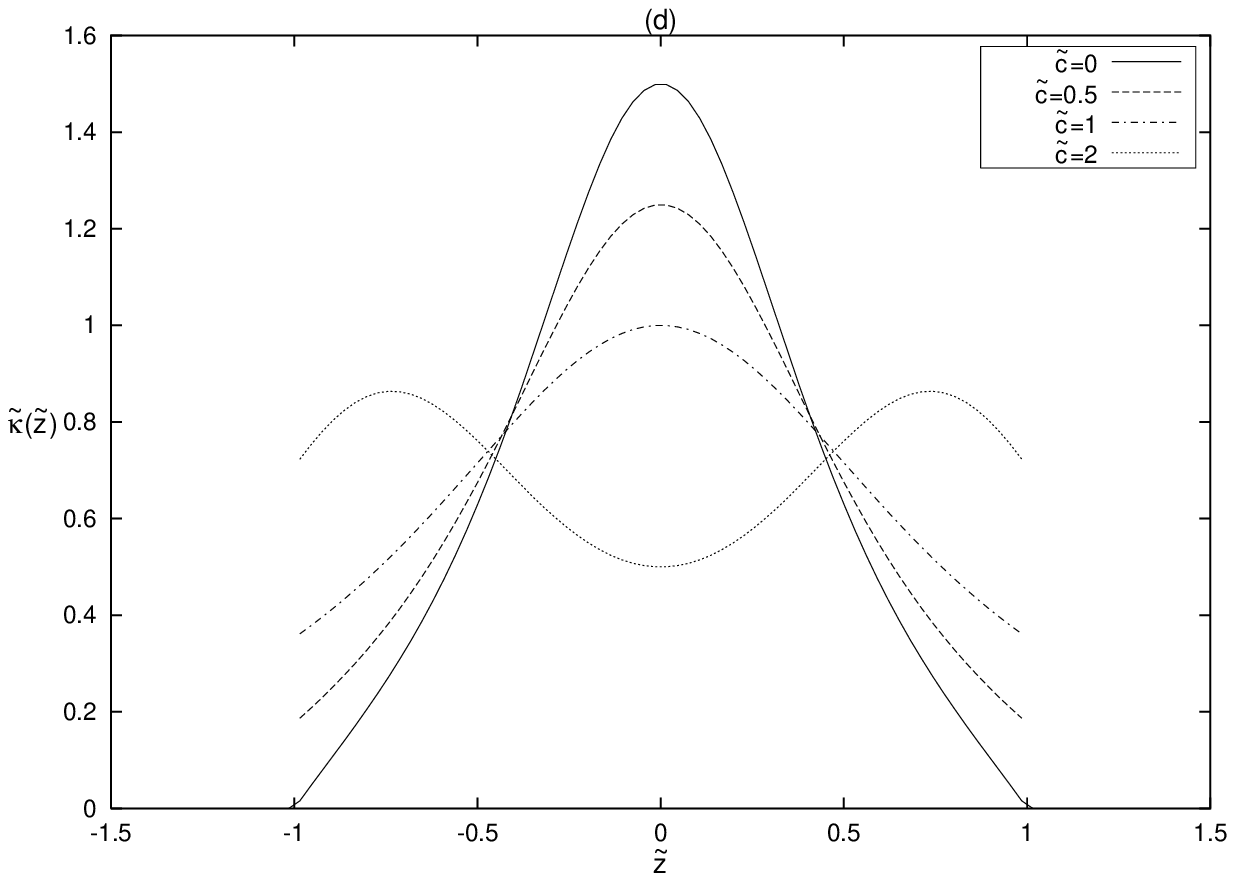}
\caption{(a) The function $\tilde{h}(\tilde{z})$ and (b) its first and (c) second derivatives with
$n=1$ for $\tilde{c}=0$, $0.5$, $1$ and $2$. (d) The curvature $\tilde{\kappa}(\tilde{z})$
for the same parameters.} \label{fig_2}
\end{figure}

\begin{figure}
\centering
\includegraphics[scale=0.75]{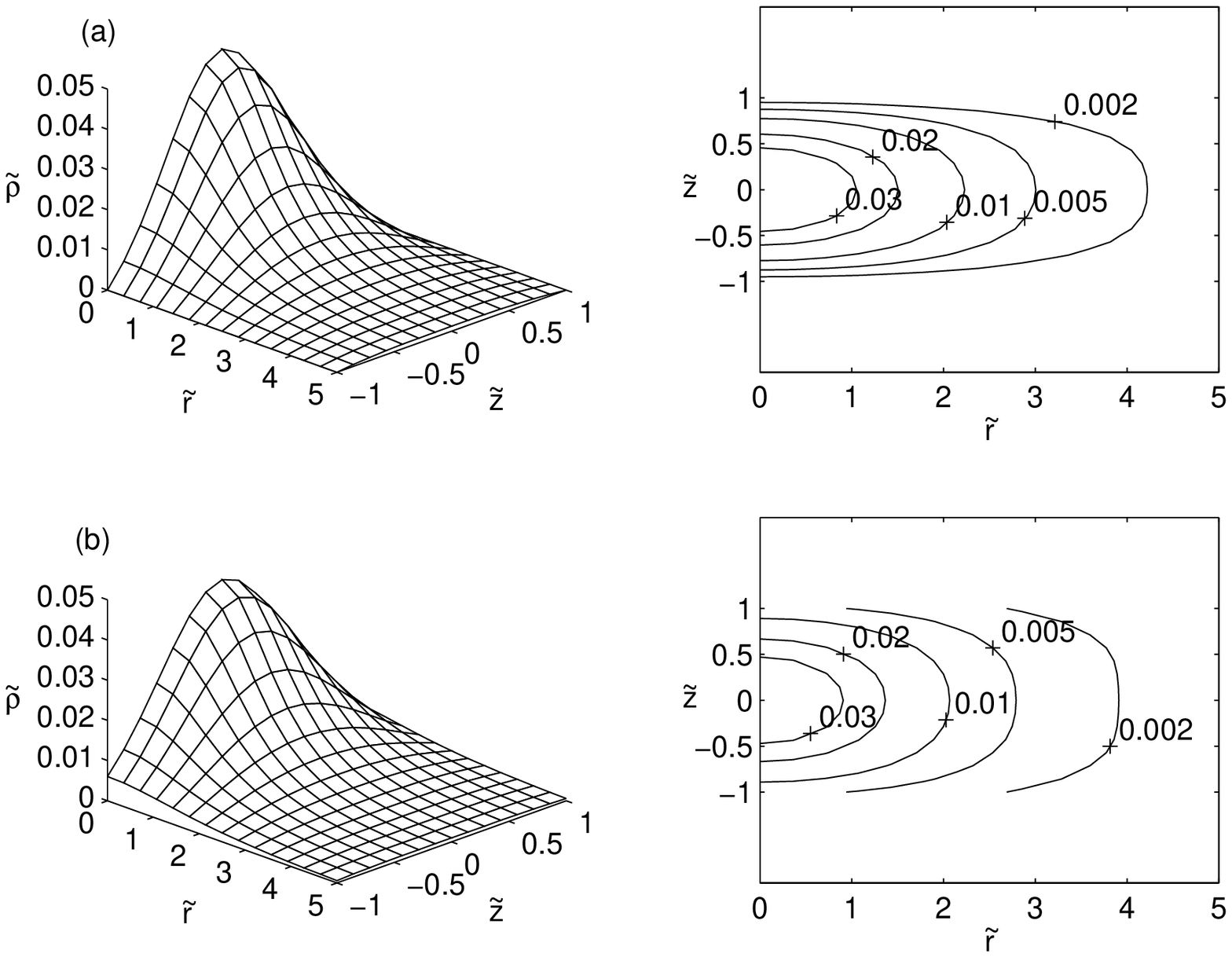}
\caption{The mass density Eq.\ (\ref{eq_rho_n2}) for a Newtonian thick disk with parameters
$\tilde{m}=1$, $\tilde{b}=2$, $n=1$ and (a) $\tilde{c}=0$, (b) $\tilde{c}=0.5$. Some levels
curves of the density are displayed on the right graphs.} \label{fig_3}
\end{figure}

\begin{figure}
\centering
\includegraphics[scale=0.75]{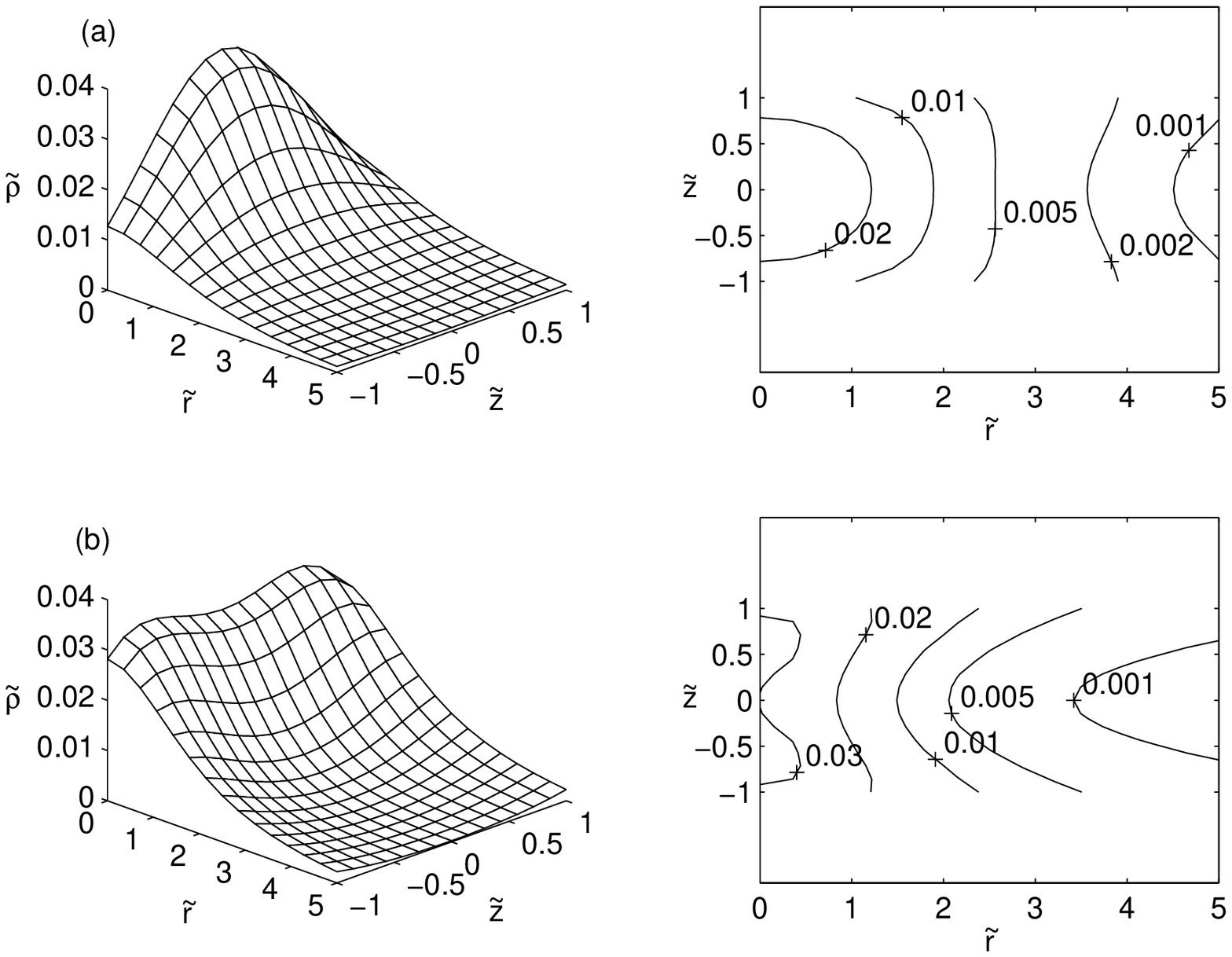}
\caption{The mass density Eq.\ (\ref{eq_rho_n2}) for a Newtonian thick disk with parameters
$\tilde{m}=1$, $\tilde{b}=2$, $n=1$ and (a) $\tilde{c}=1$, (b) $\tilde{c}=2$. Some levels
curves of the density are displayed on the right graphs.} \label{fig_4}
\end{figure}

\begin{figure}
\centering
\includegraphics[scale=0.55]{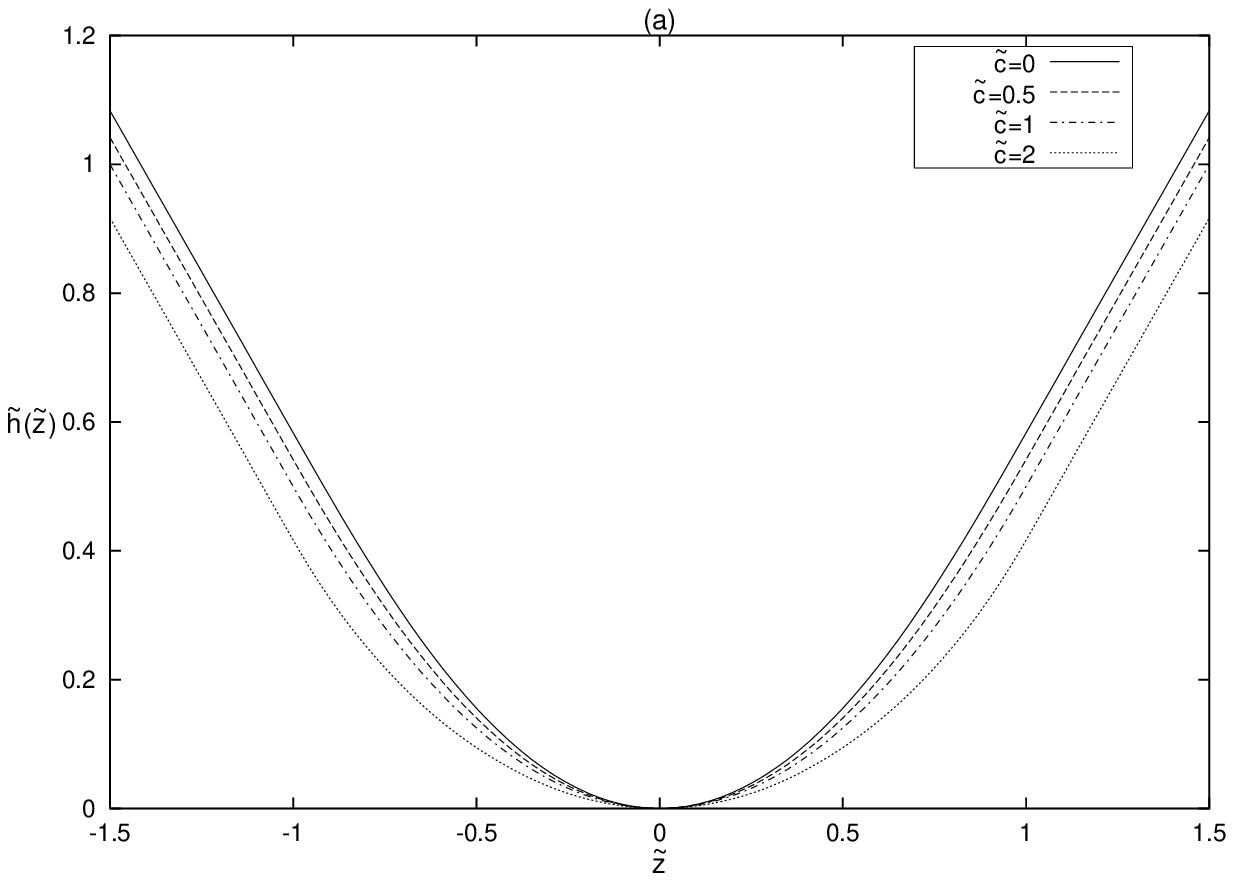}%
\includegraphics[scale=0.55]{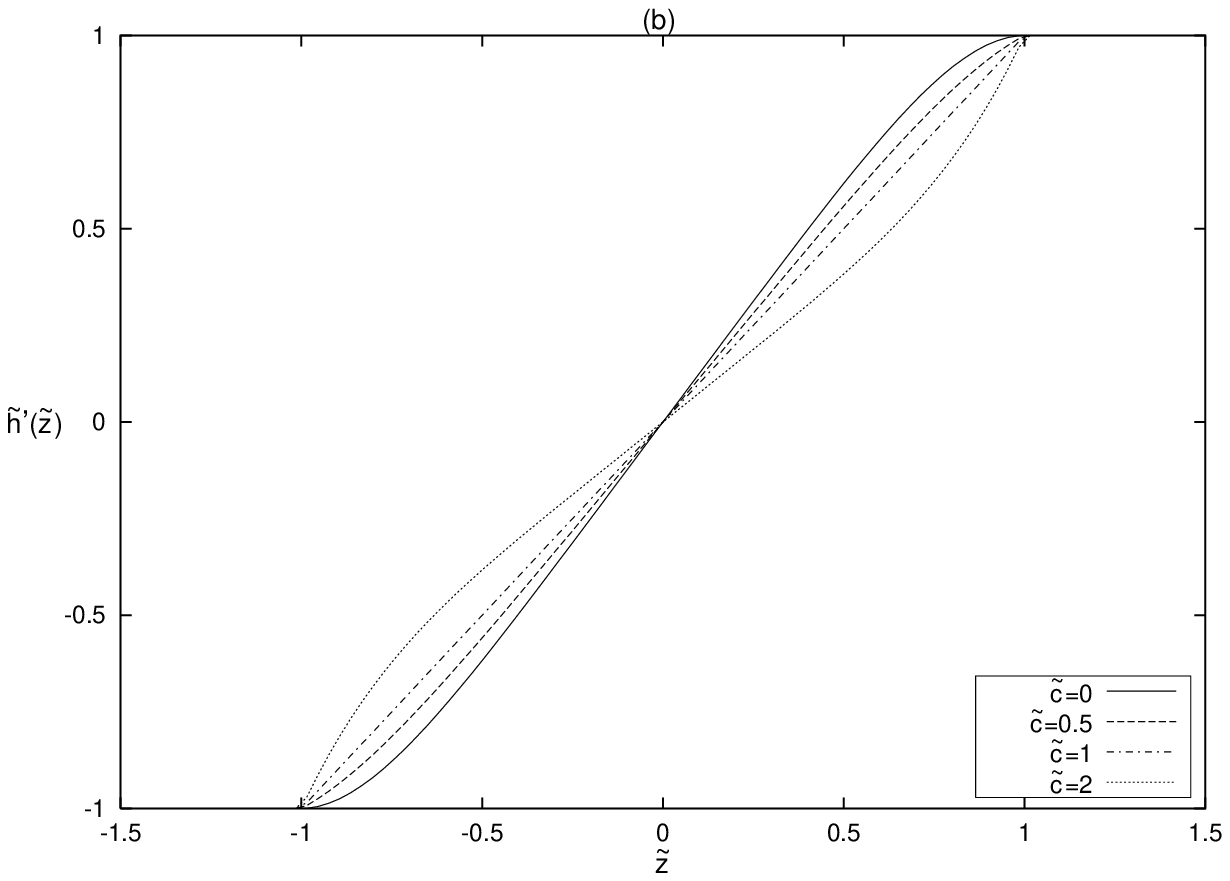}\\
\includegraphics[scale=0.55]{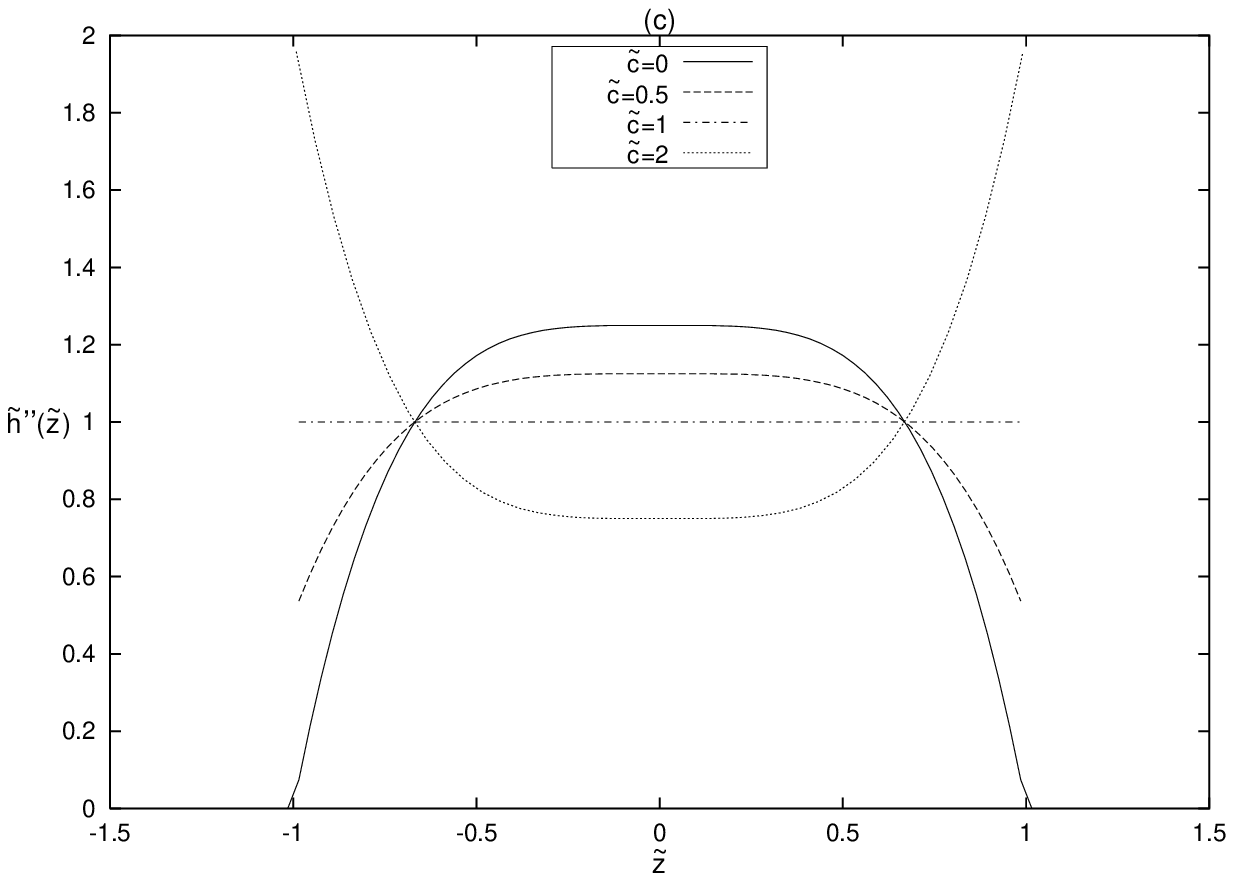}%
\includegraphics[scale=0.55]{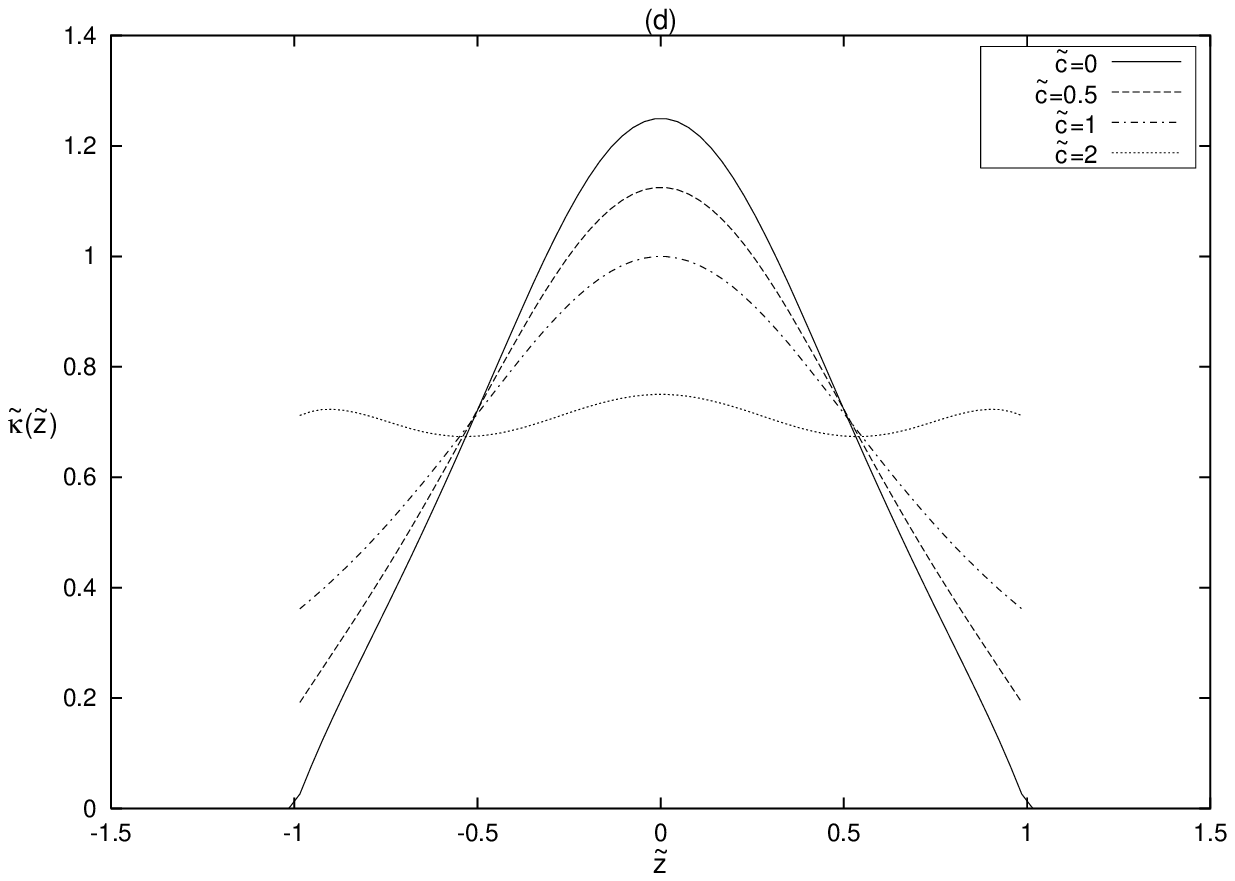}
\caption{(a) The function $\tilde{h}(\tilde{z})$ and its (b) first and (c) second derivatives with 
$n=2$ for $\tilde{c}=0$, $0.5$, $1$ and $2$. (d) The curvature $\tilde{\kappa}(\tilde{z})$
for the same parameters.} \label{fig_5}
\end{figure}

\begin{figure}
\centering
\includegraphics[scale=0.75]{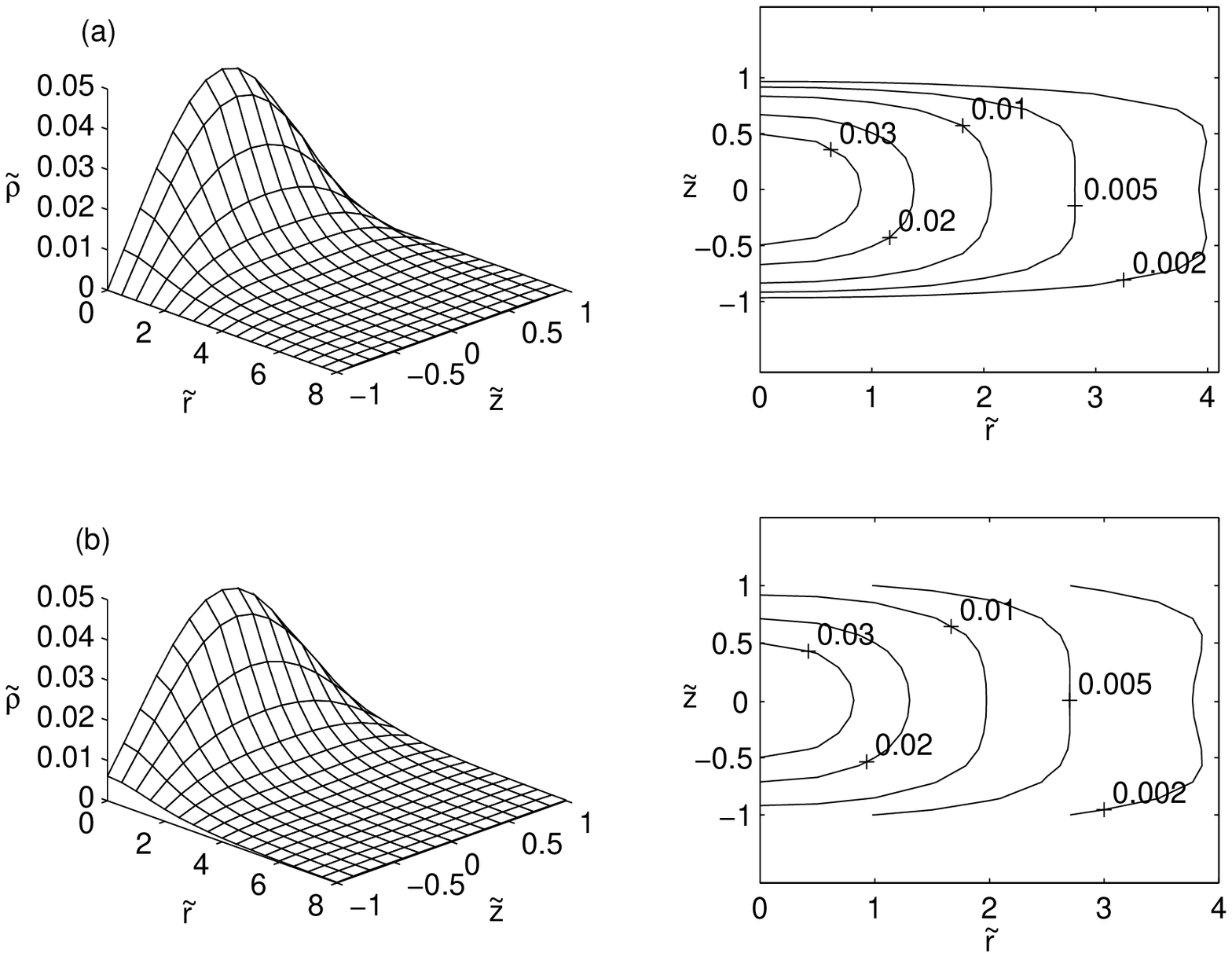}
\caption{The mass density Eq.\ (\ref{eq_rho_n2}) for a Newtonian thick disk with parameters
$\tilde{m}=1$, $\tilde{b}=2$, $n=2$ and (a) $\tilde{c}=0$, (b) $\tilde{c}=0.5$. Some levels
curves of the density are displayed on the right graphs.} \label{fig_6}
\end{figure}

\begin{figure}
\centering
\includegraphics[scale=0.75]{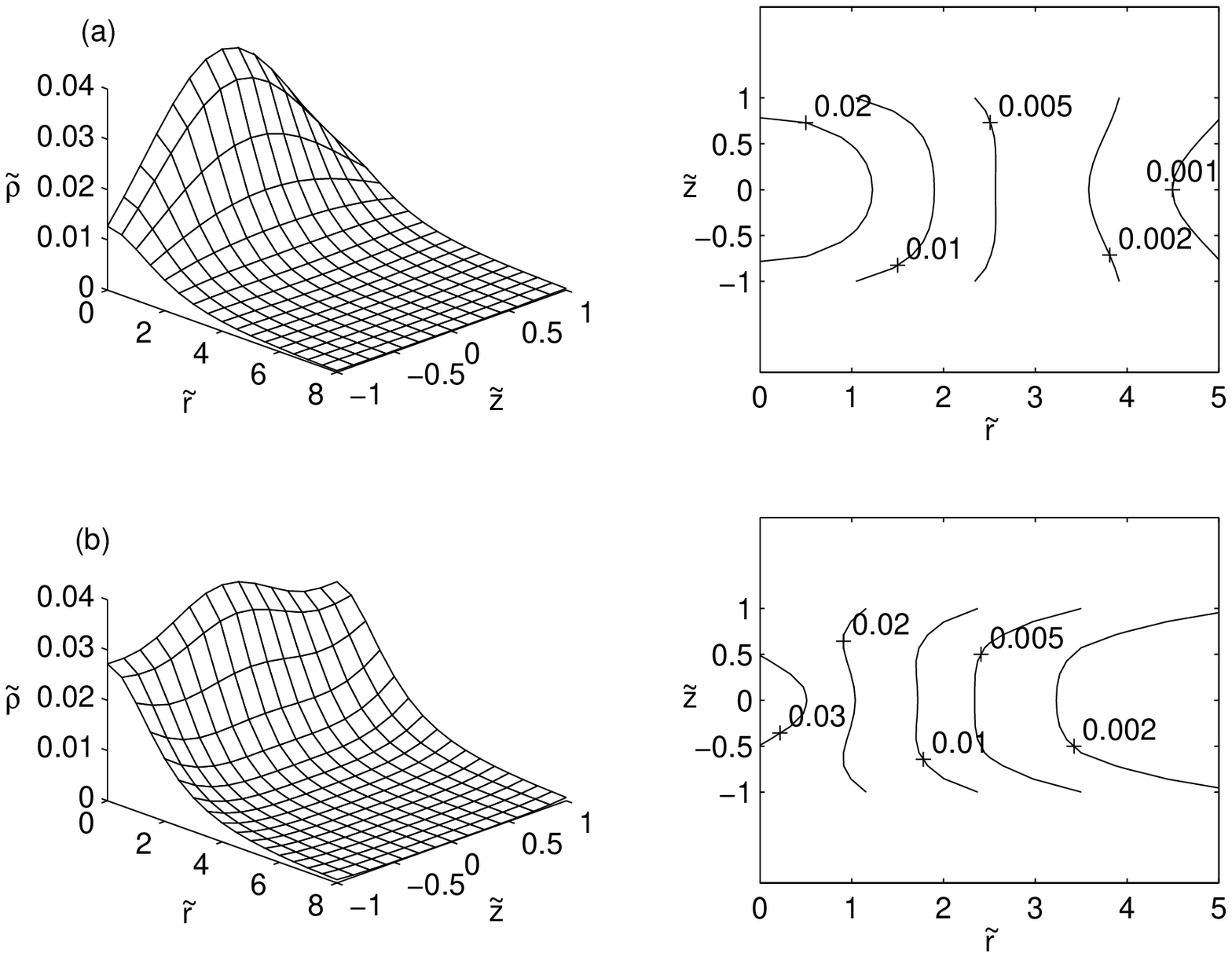}
\caption{The mass density Eq.\ (\ref{eq_rho_n2}) for a Newtonian thick disk with parameters
$\tilde{m}=1$, $\tilde{b}=2$, $n=2$ and (a) $\tilde{c}=1$, (b) $\tilde{c}=2$. Some levels
curves of the density are displayed on the right graphs.} \label{fig_7}
\end{figure}

In Fig.\ \ref{fig_2} we plot the curves of $\tilde{h}(\tilde{z})$ and its first and second 
derivatives with $n=1$ for $\tilde{c}=0$, $0.5$, $1$ and $2$. We also plot the
curvature $\tilde{\kappa}$ of $\tilde{h}$, calculated from the expression
\begin{equation} \label{eq_curv}
\tilde{\kappa}=\frac{|\tilde{h}^{\prime \prime}|}{(1+\tilde{h}^{\prime 2})^{3/2}} \mbox{.}
\end{equation} 
Fig.\ \ref{fig_3} and Fig.\ \ref{fig_4} show the mass density for a Newtonian thick disk calculated
with the function $\tilde{h}(\tilde{z})$ depicted in Fig.\ \ref{fig_2}(a) and with parameters
$\tilde{m}=1$ and $\tilde{b}=2$. In Fig.\ \ref{fig_5} we plot again $\tilde{h}(\tilde{z})$
with $n=2$ and the same values for $\tilde{c}$ as in Fig.\ \ref{fig_2}; and the mass density 
for a Newtonian disk using this function with parameters $\tilde{m}=1$ and $\tilde{b}=2$ is shown in
Fig.\ \ref{fig_6} and Fig.\ \ref{fig_7}. We note that the mass distribution along $\tilde{r}=0$ has a similar shape
to the curves for the curvature $\tilde{\kappa}$. Above some value of the jump $\tilde{c}$, the 
maximum of the mass density at $(\tilde{r},\tilde{z})=(0,0)$ becomes a local minimum point 
and two other maximum points appear. This may be interpreted as a split from one into two 
disklike distributions of matter, a configuration that does not seem to be physically reasonable. 
Thus we restrict the interval of values for $\tilde{c}$ such that the mass density has only a 
central maximum. The critical points of the curvature function are given by
\begin{equation}
\tilde{h}^{\prime \prime \prime}(1+\tilde{h}^{\prime 2})-3h^{\prime}\tilde{h}^{\prime \prime 2}=0 \mbox{.}
\end{equation}
For $n=1$, this condition leads to $\tilde{z}=0$ and the roots of
\begin{equation} \label{eq_pol1}
 7(\tilde{c}-1)^3\tilde{z}^6+11(3-\tilde{c})(\tilde{c}-1)^2\tilde{z}^4+
5(3-\tilde{c})^2(\tilde{c}-1)\tilde{z}^2-8(\tilde{c}-1)+(3-\tilde{c})^3=0 \mbox{.}
\end{equation}
To have $\tilde{z}=0$ as only critical point, Eq.\ (\ref{eq_pol1}) should not have real roots. A graphical 
analysis shows that this
happens if $\tilde{c} \lessapprox 1.46$. For $n=2$, the polynomial equation in question is
\begin{equation}
 55(\tilde{c}-1)^3\tilde{z}^{12}+65(5-\tilde{c})(\tilde{c}-1)^2\tilde{z}^8+
13(5-\tilde{c})^2(\tilde{c}-1)\tilde{z}^4-320(\tilde{c}-1)\tilde{z}^2+3(5-\tilde{c})^3=0 \mbox{,}
\end{equation}
which does not have real roots if $\tilde{c} \lessapprox 1.81$. 

The level curves in Fig.\ \ref{fig_4} and in Fig.\ \ref{fig_6}-\ref{fig_7} also indicate some
undesirable features of the disks: some isodensity curves show that the density decreases
towards the disk's center. This happens even for $n=1$, $\tilde{c}=1$ in Fig.\ \ref{fig_4}(a) and
$n=2$, $\tilde{c}=0$ in Fig.\ \ref{fig_6}(a). Thus, in practice, the ranges for the parameter $\tilde{c}$ that allow physically
acceptable disks is even more restricted than those stated above. Using graphical analysis, we find 
that large values for the ``cut'' parameter $\tilde{b}$ allow a larger ``good'' range for the parameter
$\tilde{c}$, but in the next sections when the procedure will be applied in general relativity, large 
values for $\tilde{b}$ will also mean less relativistic disks.

\section{Thick Disks from the Schwarzschild Metric in Isotropic Coordinates} \label{sec_iso}

In cylindrical coordinates $(t,r,z,\varphi)$ the isotropic metric takes the form
\begin{equation} \label{eq_metric_iso}
\mathrm{d}s^2=-e^{2\nu}\mathrm{d}t^2+e^{2\mu}(\mathrm{d}r^2+\mathrm{d}z^2+
r^2\mathrm{d}\varphi^2) \mbox{,}
\end{equation}
where the functions $\nu$ and $\mu$ depend on $r$ and $z$. The Schwarzschild solution 
for metric Eq.\ (\ref{eq_metric_iso}) can be expressed as
\begin{subequations}
\begin{align}
\nu &= \ln \left[ \frac{2R-m}{2R+m} \right] \mbox{,} \label{eq_sch_iso1} \\
\mu &= 2 \ln \left[ 1+\frac{m}{2R} \right] \mbox{,} \label{eq_sch_iso2}
\end{align}
\end{subequations}
where $m>0$ and $R^2=r^2+z^2$. The ``displace, cut, fill and reflect'' method applied 
to the above solution is equivalent to put $R^2=r^2+(h+b)^2$ where $b>0$ and $h(z)$ given 
by Eq.\ (\ref{eq_h}). The continuity of the metric function and its derivatives on $z=\pm a$ is 
satisfied by the continuity of $h(z)$ and $h^{\prime}(z)$ on $z=\pm a$. For $|z|>a$, 
Eq.\ (\ref{eq_metric_iso}) satisfies the vacuum Einstein equations. For $|z|<a$, the components 
of the energy-momentum tensor of the disk are computed from the Einstein equations 
\begin{equation} \label{eq_einstein}
T_{ab}=R_{ab}-\frac{1}{2}g_{ab}R \mbox{,}
\end{equation}
where we use units such that $c=8\pi G=1$. Defining the orthonormal tetrad
$\{V^a,W^a,Y^a,Z^a\}$, with
\begin{subequations}
\begin{align}
V^a &=e^{-\nu}(1,0,0,0) \mbox{,}\\
W^a &=e^{-\mu}(0,1,0,0) \mbox{,}\\
Y^a &=e^{-\mu}(0,0,1,0) \mbox{,}\\
Z^a &=\frac{e^{-\mu}}{r}(0,0,0,1) \mbox{,}
\end{align}
\end{subequations}
and using Eq.\ (\ref{eq_einstein}), the energy-momentum tensor can be written as
\begin{equation} \label{eq_emt_iso}
T_{ab}=\sigma V_aV_b+p_r W_aW_b+p_z Y_aY_b+p_{\varphi}Z_aZ_b \mbox{,}
\end{equation}
where $\sigma=-T^t_t$ is the energy density, $p_r=T^r_r$ is the radial stress, 
which is equal to the azimuthal stress $p_{\varphi}=T^{\varphi}_{\varphi}$, and 
$p_z=T^z_z$ is the vertical stress. The effective Newtonian density is given by 
$\rho=\sigma+p_r+p_z+p_{\varphi}$. To satisfy the strong energy condition we must have 
$\rho \geq 0$, the weak energy condition requires $\sigma \geq 0$ and the dominant 
energy condition requires $|p_r/\sigma| \leq 1$, $|p_z/\sigma| \leq 1$ and 
$|p_{\varphi}/\sigma| \leq 1$. Using Eq.\ (\ref{eq_sch_iso1})-(\ref{eq_sch_iso2}), 
we obtain \cite{errata1}
\begin{subequations}
\begin{align}
\tilde{\sigma} &=\frac{64\tilde{m}}{(\tilde{m}+2\tilde{R})^5}\left[ 3(\tilde{h}+\tilde{b})^2
(1-\tilde{h}^{\prime 2})+\tilde{R}^2[\tilde{h}^{\prime 2}-1+\tilde{h}^{\prime \prime}
(\tilde{h}+\tilde{b})] \right] \mbox{,} \label{eq_sigma_iso} \\
\tilde{\rho} &=\frac{128\tilde{m}\tilde{R}}{(\tilde{m}+2\tilde{R})^5(-\tilde{m}+2\tilde{R})}
\left[3(\tilde{h}+\tilde{b})^2(1-\tilde{h}^{\prime 2})+\tilde{R}^2[\tilde{h}^{\prime 2}-
1+\tilde{h}^{\prime \prime}(\tilde{h}+\tilde{b})] \right] \mbox{,} \label{eq_rho_iso} \\
\tilde{p}_r &= \tilde{p}_{\varphi}=\frac{32\tilde{m}^2}{(\tilde{m}+2\tilde{R})^5(-\tilde{m}+2\tilde{R})}
\left[2(\tilde{h}+\tilde{b})^2(1-\tilde{h}^{\prime 2})+ \tilde{R}^2[\tilde{h}^{\prime 2}-
1+\tilde{h}^{\prime \prime}(\tilde{h}+\tilde{b})] \right] \mbox{,} \label{eq_pr_iso} \\
\tilde{p}_z &=\frac{64\tilde{m}^2(1-\tilde{h}^{\prime 2})(\tilde{h}+\tilde{b})^2}{(\tilde{m}+2\tilde{R})^5
(-\tilde{m}+2\tilde{R})} \mbox{,} \label{eq_pz_iso}
\end{align}
\end{subequations}
where $\tilde{\sigma}=a^2\sigma$, $\tilde{\rho}=a^2\rho$, $\tilde{p}_r=a^2p_r$,
$\tilde{p}_z=a^2p_z$, and the other dimensionless variables were defined in Sec.\ \ref{sec_newt}.

From Eq.\ (\ref{eq_sigma_iso})-(\ref{eq_pr_iso}) we can see that if condition (\ref{eq_cond_n}) is
satisfied, we have $\tilde{\sigma} \geq 0$, $\tilde{\rho}\geq0$ and
$\tilde{p}_r=\tilde{p}_{\varphi} \geq 0$ (pressures). Thus $\tilde{b} \geq 2n/(2n+1-\tilde{c})$ 
and $0 \leq \tilde{c} \leq 2n+1$. We also have vertical pressure.
To ensure nonsingular behaviour of the expressions, we impose that $0<\tilde{m}<2\tilde{b}$. 
\begin{figure}
\centering
\includegraphics[scale=0.75]{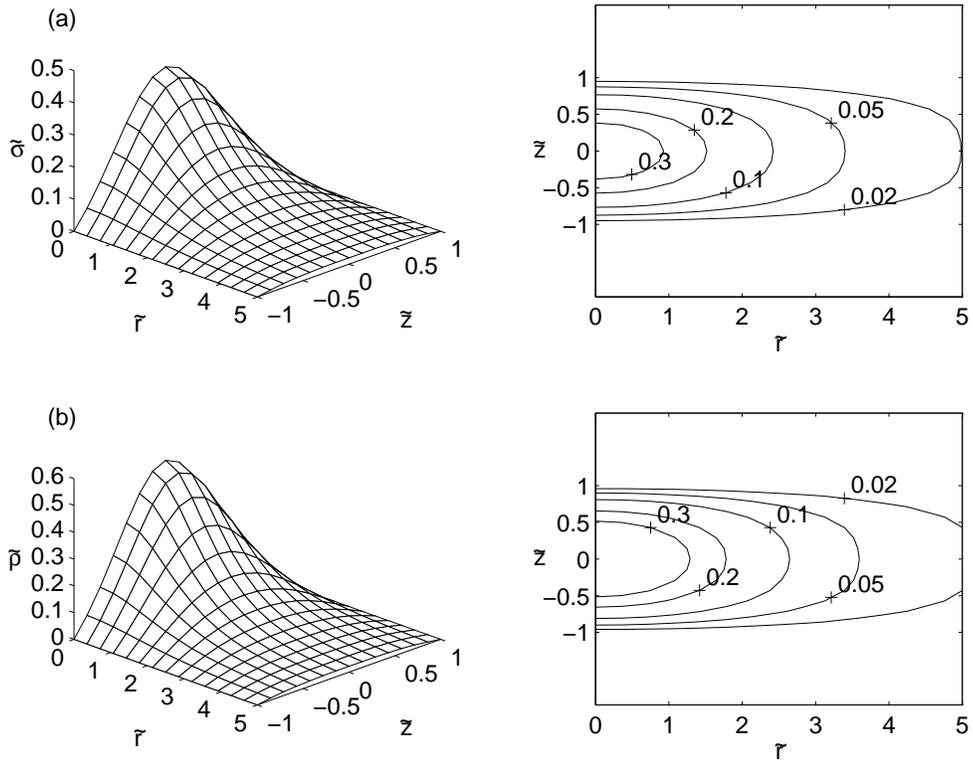}
\caption{(a) Energy density, (b) effective Newtonian density for a thick disk in isotropic coordinates.
Parameters: $\tilde{m}=1$, $\tilde{b}=2$, $n=1$ and $\tilde{c}=0$.} \label{fig_8}
\end{figure}

\begin{figure}
\centering
\includegraphics[scale=0.75]{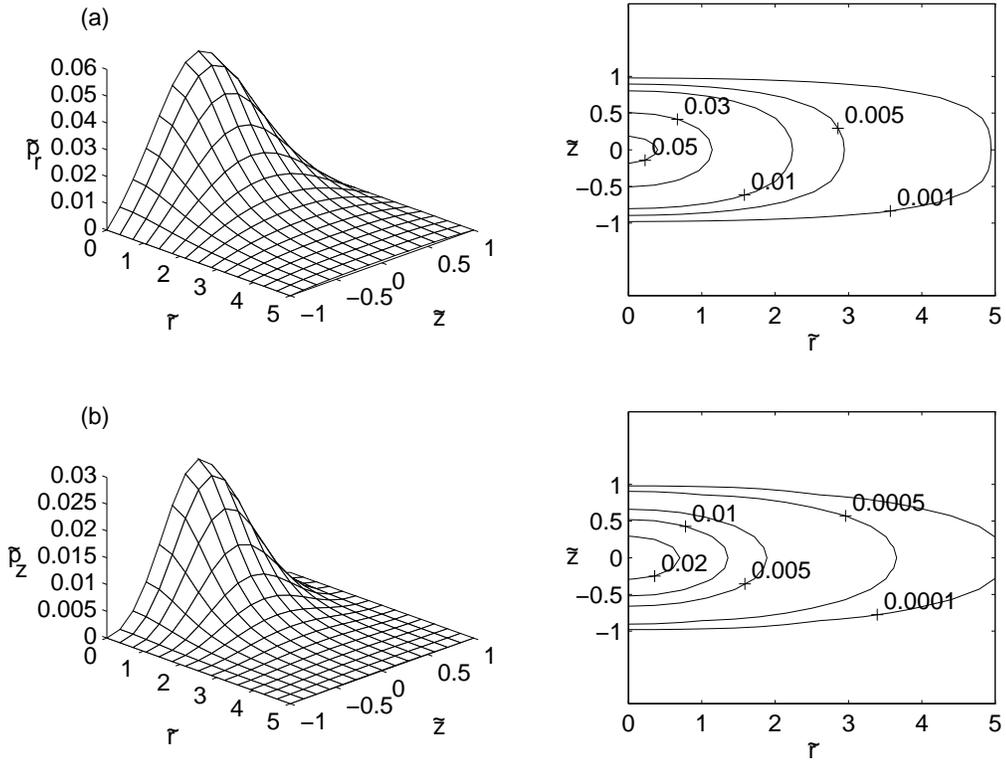}
\caption{(a) Radial and azimuthal pressures, (b) vertical pressure for a thick disk in isotropic coordinates.
Parameters: $\tilde{m}=1$, $\tilde{b}=2$, $n=1$ and $\tilde{c}=0$.} \label{fig_9}
\end{figure}
In Fig.\ \ref{fig_8} and Fig. \ref{fig_9} we plot the surfaces and level curves for the energy density, effective 
Newtonian density, radial and azimuthal pressures and vertical pressure for a
thick disk with parameters $\tilde{m}=1$, $\tilde{b}=2$, $n=1$ and $\tilde{c}=0$. The ratio between any
pressure and energy density is less than 0.15. Thus all energy conditions are satisfied. The shape of the level curves
also show that for these parameters the disk is physically acceptable.

\section{Thick Disks from the Schwarzschild Metric in Weyl Coordinates} \label{sec_weyl}
The general metric for a static axially symmetric spacetime in Weyl's canonical coordinates 
$(t,r,z,\varphi)$ is given by
\begin{equation} \label{eq_metric_w}
\mathrm{d}s^2=-e^{2\Phi}\mathrm{d}t^2+e^{-2\Phi}\left[ e^{2\Lambda}(\mathrm{d}r^2+\mathrm{d}z^2)+
r^2\mathrm{d}\varphi^2 \right] \mbox{,}
\end{equation}
where $\Phi$ and $\Lambda$ are functions of $r$ and $z$. The Einstein vacuum equations 
for this metric reduce to the Weyl equations \cite{Weyl1,Weyl2}
\begin{subequations}
\begin{align}
& \Phi_{,rr}+\frac{\Phi_r}{r}+\Phi_{,zz}=0 \mbox{,} \label{eq_weyl1}\\
& \Lambda_r=r(\Phi_r^2-\Phi_z^2) \mbox{,} \label{eq_weyl2}\\
& \Lambda_z=2r \Phi_r \Phi_z \mbox{.} \label{eq_weyl3}
\end{align}
\end{subequations}
In these coordinates, Schwarzschild solution assumes the form \cite{Weyl1}
\begin{subequations}
\begin{align}
\Phi &=\frac{1}{2} \ln \left[ \frac{R_1+R_2-2m}{R_1+R_2+2m} \right] \mbox{,} \label{eq_sch_weyl1}\\
\Lambda &=\frac{1}{2} \ln \left[ \frac{(R_1+R_2)^2-4m^2}{4R_1R_2} \right] \mbox{,}  \label{eq_sch_weyl2}
\end{align}
\end{subequations}
where $R_1^2=r^2+(m+z)^2$, $R_2^2=r^2+(-m+z)^2$ and $m>0$. By applying the 
transformation $z\rightarrow h(z)+b$ on  Eq.\ (\ref{eq_sch_weyl1})-(\ref{eq_sch_weyl2}), 
using Eq.\ (\ref{eq_weyl1})-(\ref{eq_weyl3}) and Einstein equations 
Eq.\ (\ref{eq_einstein}), the energy-momentum tensor can be written as 
\begin{equation} \label{eq_emt_weyl}
T_{ab}=\epsilon V_aV_b+p_r W_aW_b+p_z Y_aY_b+p_{\varphi}Z_aZ_b \mbox{,}
\end{equation}
whith the orthonormal tetrad $\{V^a,W^a,Y^a,Z^a\}$ defined as
\begin{subequations}
\begin{align}
V^a &=e^{-\Phi}(1,0,0,0) \mbox{,}\\
W^a &=e^{\Phi-\Lambda}(0,1,0,0) \mbox{,}\\
Y^a &=e^{\Phi-\Lambda}(0,0,1,0) \mbox{,}\\
Z^a &=\frac{e^{\Phi}}{r}(0,0,0,1) \mbox{.}
\end{align}
\end{subequations}
In Eq.\ (\ref{eq_emt_weyl}) $\epsilon=-T^t_t$ is the energy density and $p_r=T^r_r$, $p_z=T^z_z=-T^r_r$, 
$p_{\varphi}=T^{\varphi}_{\varphi}$ are respectively, the radial, vertical and azimuthal stresses. The effective 
Newtonian density is given by $\rho=\epsilon+p_{\varphi}$. The explicit expressions are
\begin{subequations}
\begin{align}
\tilde{\rho} &=\frac{e^{2(\Phi-\Lambda)}}{\tilde{R}_1^3\tilde{R}_2^3} \left\{ 
\tilde{h}^{\prime \prime}(\tilde{R}_1-\tilde{R}_2)\tilde{R}_1^2\tilde{R}_2^2+
(1-\tilde{h}^{\prime 2})\left[ (\tilde{h}+\tilde{b})(\tilde{R}_1^3-\tilde{R}_2^3)-\tilde{m}
(\tilde{R}_1^3+\tilde{R}_2^3)\right] \right\} \mbox{,} \label{eq_rho_weyl}\\
\tilde{\epsilon} &= \tilde{\rho} \left[ 1-\frac{2\tilde{m}\tilde{r}^2(\tilde{R}_1+
\tilde{R}_2)}{\tilde{R}_1\tilde{R}_2[(\tilde{R}_1+\tilde{R}_2)^2-4\tilde{m}^2]} \right]
+\frac{e^{2(\Phi-\Lambda)}(1-\tilde{h}^{\prime 2})(\tilde{R}_1-\tilde{R}_2)}{4\tilde{R}_1^4\tilde{R}_2^4} \times \notag \\
& \left[ (\tilde{R}_1-\tilde{R}_2)\tilde{R}_1^2\tilde{R}_2^2-2\tilde{r}^2
(\tilde{R}_1^3-\tilde{R}_2^3) \right] \mbox{,} \label{eq_eps_weyl}\\
\tilde{p}_{\varphi} &= \frac{2\tilde{m}\tilde{\rho}\tilde{r}^2(\tilde{R}_1+\tilde{R}_2)}
{\tilde{R}_1\tilde{R}_2[(\tilde{R}_1+\tilde{R}_2)^2-4\tilde{m}^2]}+
\frac{e^{2(\Phi-\Lambda)}(\tilde{h}^{\prime 2}-1)(\tilde{R}_1-\tilde{R}_2)}{4\tilde{R}_1^4\tilde{R}_2^4} \times \notag \\
& \left[ (\tilde{R}_1-\tilde{R}_2)\tilde{R}_1^2\tilde{R}_2^2-2\tilde{r}^2
(\tilde{R}_1^3-\tilde{R}_2^3) \right] \mbox{,} \label{eq_pfi_weyl}\\
\tilde{p}_r &= \frac{4\tilde{m}^2e^{2(\Phi-\Lambda)}(\tilde{h}^{\prime 2}-1)(\tilde{h}+\tilde{b})^2}
{\tilde{R}_1^2\tilde{R}_2^2(\tilde{R}_1+\tilde{R}_2)^2} \mbox{,} \label{eq_pr_weyl}\\
\tilde{p}_z &= \frac{4\tilde{m}^2e^{2(\Phi-\Lambda)}(1-\tilde{h}^{\prime 2})(\tilde{h}+\tilde{b})^2}
{\tilde{R}_1^2\tilde{R}_2^2(\tilde{R}_1+\tilde{R}_2)^2} \mbox{,} \label{eq_pz_weyl}
\end{align}
\end{subequations}
From Eq.\ (\ref{eq_pr_weyl}) and Eq.\ (\ref{eq_pz_weyl}) we have vertical pressures and radial tensions.
From Eq.\ (\ref{eq_eps_weyl}) the condition $\tilde{\epsilon}\geq 0$ on $|\tilde{z}|=1$ gives
\begin{equation}
\frac{2\tilde{m}\tilde{r}^2(\tilde{R}_1+
\tilde{R}_2)}{\tilde{R}_1\tilde{R}_2[(\tilde{R}_1+\tilde{R}_2)^2-4\tilde{m}^2]} \leq 1 \mbox{.}
\end{equation}
Since $\tilde{R}_1>\tilde{r}$, $\tilde{R}_2>\tilde{r}$, and $\tilde{R}_1+\tilde{R}_2\geq 2\tilde{b}$, 
we have 
\begin{equation}
\frac{\tilde{m}\tilde{b}}{\tilde{b}^2-\tilde{m}^2} \leq 1 \mbox{,}
\end{equation}
which is equivalent to $\tilde{m} \leq (\sqrt{5}-1)\tilde{b}/2$ \cite{errata2}. It is not easy to obtain other constraints
over the parameters $\tilde{m}$ and $\tilde{b}$ in order to satisfy all energy conditions. The analysis 
is better done graphically. 
\begin{figure}
\centering
\includegraphics[scale=0.75]{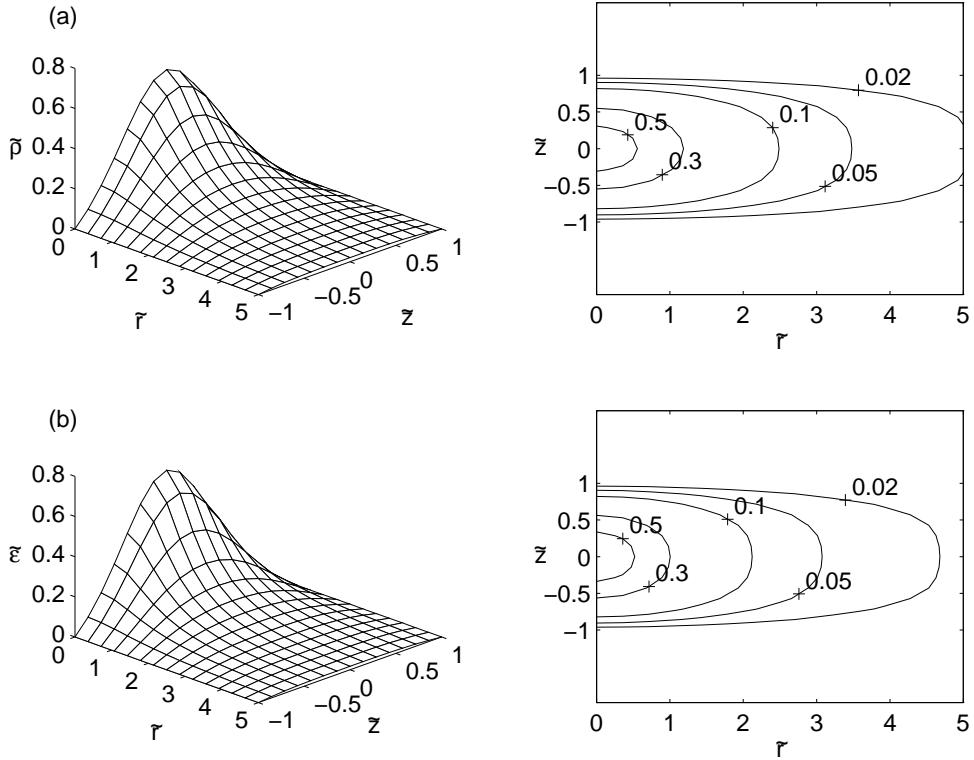}
\caption{(a) Effective Newtonian density, (b) energy density for a thick disk in Weyl coordinates. 
Parameters: $\tilde{m}=1$, $\tilde{b}=2$, $n=1$ and $\tilde{c}=0$.} \label{fig_10}
\end{figure}

\begin{figure}
\centering
\includegraphics[scale=0.75]{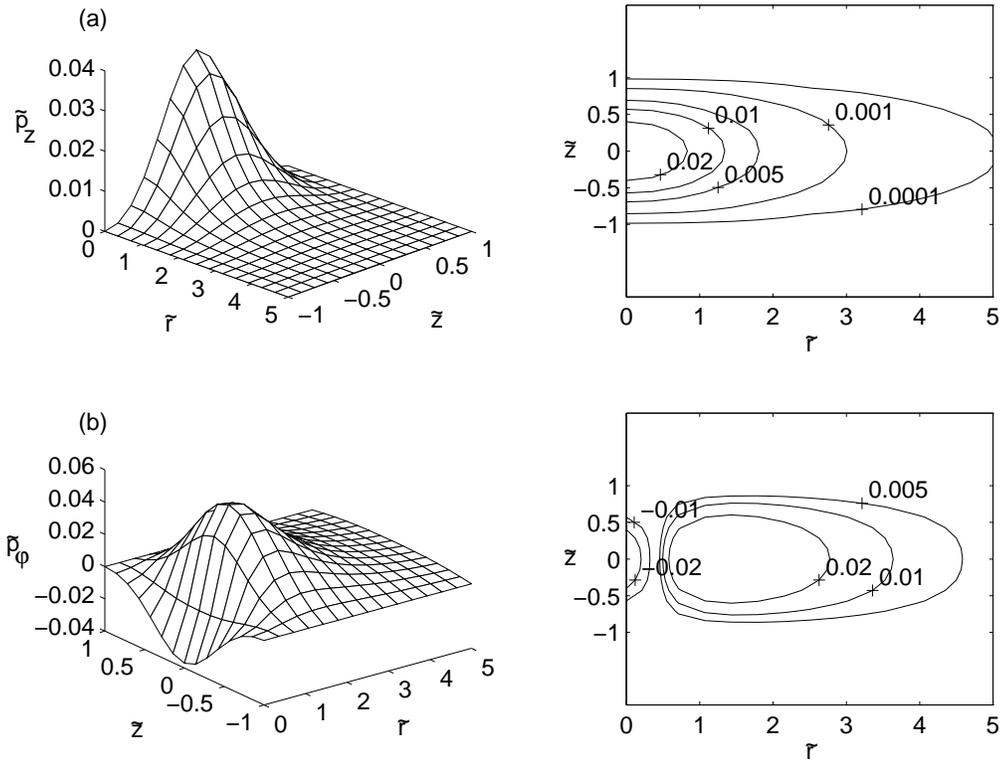}
\caption{(a) Vertical pressures, (b) azimuthal stresses for a thick disk in Weyl coordinates. 
Parameters: $\tilde{m}=1$, $\tilde{b}=2$, $n=1$ and $\tilde{c}=0$.} \label{fig_11}
\end{figure}
In Fig.\ \ref{fig_10} and Fig.\ \ref{fig_11} we plot the surfaces and level curves for the effective 
Newtonian density, energy density, vertical pressure and the azimuthal stresses for a thick disk with parameters
$\tilde{m}=1$, $\tilde{b}=2$, $n=1$, and $\tilde{c}=0$.
For these values we have $\tilde{\rho} \geq 0$ and $\tilde{\epsilon} \geq 0$ everywhere.
The ratio between any pressure or tension and energy density is less than 0.35,
thus the dominant energy condition is also satisfied.

\section{Thick Disks from the Schwarzschild Metric in Schwarzschild Coordinates} \label{sec_schw}
The Schwarzschild metric in Schwarzschild canonical coordinates 
$(t,r,\theta,\varphi)$ is written as
\begin{equation} \label{eq_metric_sch1}
\mathrm{d}s^2=-\left( 1-\frac{2m}{r} \right)\mathrm{d}t^2+\frac{1}{\left( 1-\frac{2m}{r} \right)}
\mathrm{d}r^2+r^2(\mathrm{d}\theta^2+\sin^2 \theta \mathrm{d}\varphi^2) \mbox{,}
\end{equation}
where $m$ is a positive constant. In cylindrical coordinates $(t,R,z,\varphi)$, 
Eq.\ (\ref{eq_metric_sch1}) can be cast as
\begin{align} \label{eq_metric_sch2}
\mathrm{d}s^2 &=-\left( 1-\frac{2m}{\mathcal{R}} \right)\mathrm{d}t^2+
\frac{\mathcal{R}^3-2mz^2}{\mathcal{R}^2(\mathcal{R}-2m)} \mathrm{d}R^2+
\frac{4mRz}{\mathcal{R}^2(\mathcal{R}-2m)} \mathrm{d}R\mathrm{d}z+ \notag \\
& \frac{\mathcal{R}^3-2mR^2}{\mathcal{R}^2(\mathcal{R}-2m)} \mathrm{d}z^2+R^2\mathrm{d}\varphi^2 \mbox{,}
\end{align}
where $\mathcal{R}^2=R^2+z^2$. The ``displace, cut, fill and reflect'' method cannot 
be directly applied to metric Eq.\ (\ref{eq_metric_sch2}), since the component 
$g_{Rz}$ is not an even function of $z$. But we can multiply it by an odd function 
of $z$, which we choose to be the first derivative of $h(z)$. So we make the 
transformations $z\rightarrow h(z)+b$ \textit{and} $g_{Rz}\rightarrow g_{Rz}h^{\prime}(z)$
in Eq.\ (\ref{eq_metric_sch2}), and use the Einstein equations Eq.\ (\ref{eq_einstein}) 
to compute the components of the disk's energy-momentum tensor. With $h(z)$ defined 
by Eq.\ (\ref{eq_h}), the metric Eq.\ (\ref{eq_metric_sch2}) satisfies the vacuum Einstein 
equations for $|z|>a$ and the metric functions together with their derivatives with 
respect to $z$ are continuous on $z=\pm a$. Note that the function multiplying $g_{Rz}$ may 
be other kind of function not necerrarily related to $h(z)$, but its possible form is also limited by 
the requirements that the generated disks are physically acceptable, as was discussed in Sec.\ \ref{sec_newt}. 
The physical variables of the disk are obtained by solving the eigenvalue problem for $T^a_b$
\begin{equation}
T^a_b\xi^b=\lambda\xi^a \mbox{.}
\end{equation}
We find that $T^{ab}$ can be put in the form
\begin{equation} \label{eq_emt_sch}
T^{ab}=\epsilon U^aU^b+p_{+}V^aV^b+p_{-}X^aX^b+p_{\varphi}W^aW^b \mbox{,}
\end{equation}
where
\begin{align}
\epsilon &=-T^t_t, \qquad U^a=\frac{1}{\sqrt{-g_{tt}}}(1,0,0,0) \mbox{,} \notag \\
p_{\pm} &=\frac{T^r_r+T^z_z}{2} \pm \frac{1}{2}\sqrt{(T^r_r-T^z_z)^2+4T^r_zT^z_r} 
\mbox{,} \notag \\
V^a &=(0,V^r,V^z,0), \qquad X^a=(0,X^r,X^z,0)  \mbox{,} \notag \\
p_{\varphi} &= T^{\varphi}_{\varphi}, \qquad W^a=\frac{1}{R}(0,0,0,1) \mbox{,}
\end{align}
and 
\begin{align}
V^r &=\frac{1}{\sqrt{g_{RR}-2g_{Rz}\Delta+g_{zz}\Delta^2}}, \quad
V^z =-\Delta V^r, \quad \Delta =\frac{T^r_r-p_{+}}{T^r_z} \mbox{,} \notag \\
X^r &=\frac{1}{\sqrt{g_{RR}-2g_{Rz}\Gamma+g_{zz}\Gamma^2}}, \quad
X^z =-\Gamma X^r, \quad \Gamma =\frac{T^r_r-p_{-}}{T^r_z} \mbox{.}
\end{align}
The effective Newtonian density reads $\rho=\epsilon+p_{+}+p_{-}+p_{\varphi}=
\epsilon+T^r_r+T^z_z+p_{\varphi}$.
\begin{figure}
\centering
\includegraphics[scale=0.75]{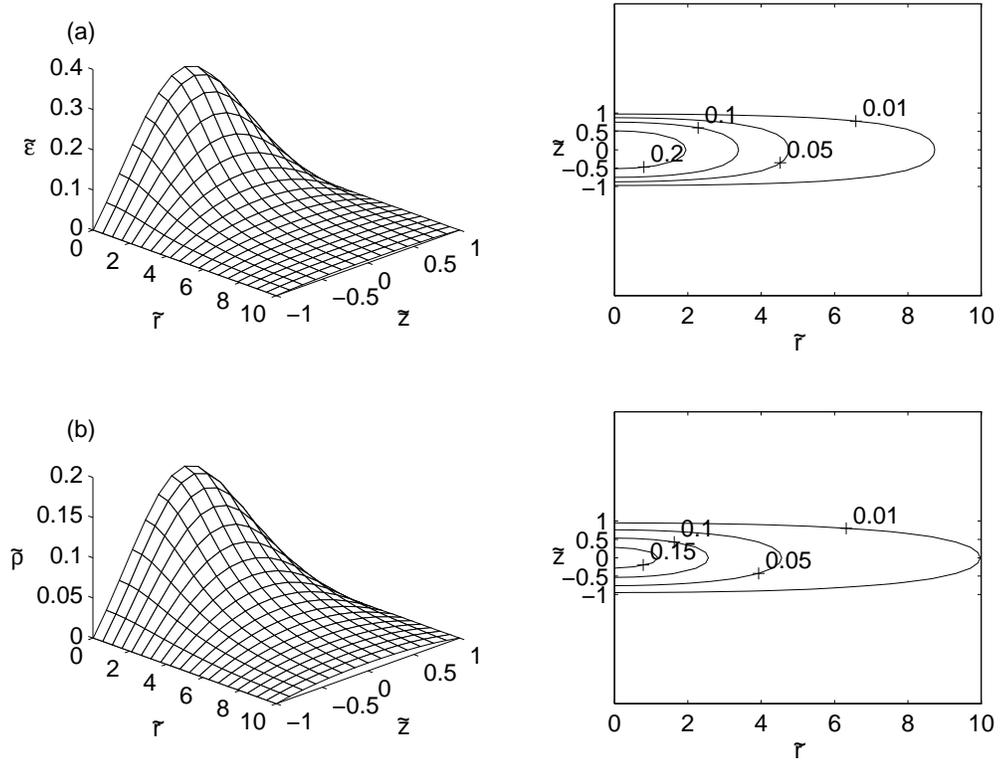}
\caption{(a) Energy density, (b) effective Newtonian density for a thick disk in Schwarzschild
coordinates. Parameters: $\tilde{m}=1$, $\tilde{b}=5$, $n=1$ and $\tilde{c}=0$.} \label{fig_12}
\end{figure}

\begin{figure}
\centering
\includegraphics[scale=0.75]{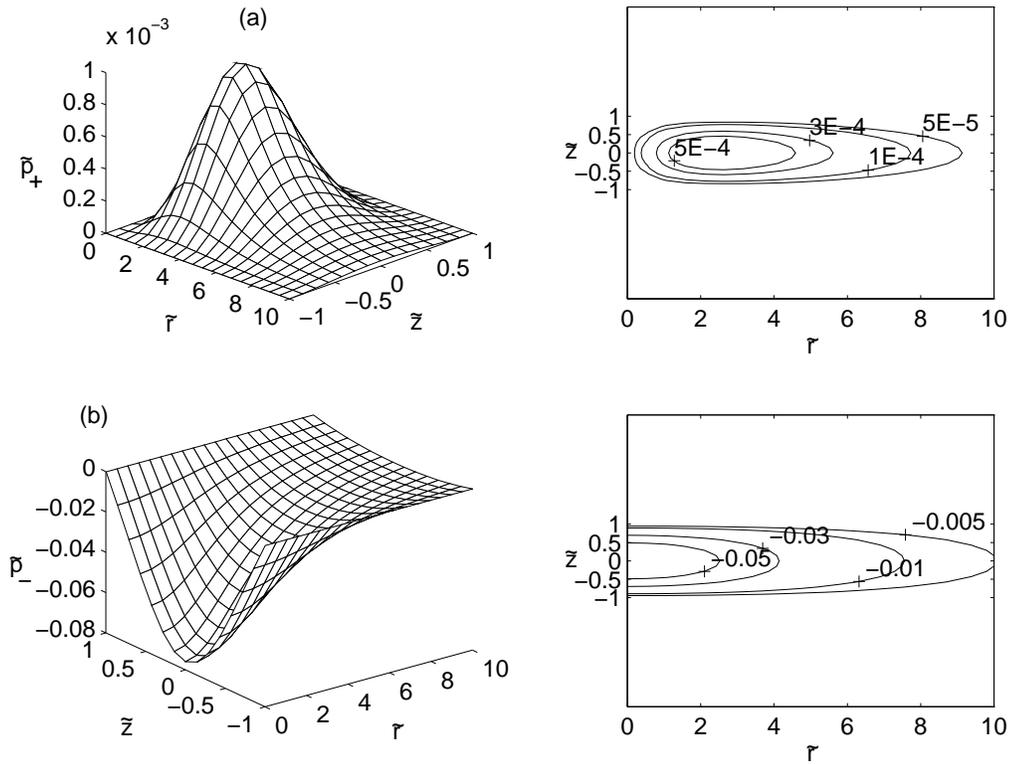}
\caption{(a) The pressure mostly vertical, (b) tension mostly radial for a thick disk in Schwarzschild
coordinates. Parameters: $\tilde{m}=1$, $\tilde{b}=5$ and $\tilde{c}=0$.} \label{fig_13}
\end{figure}

\begin{figure}
\centering
\includegraphics[scale=0.75]{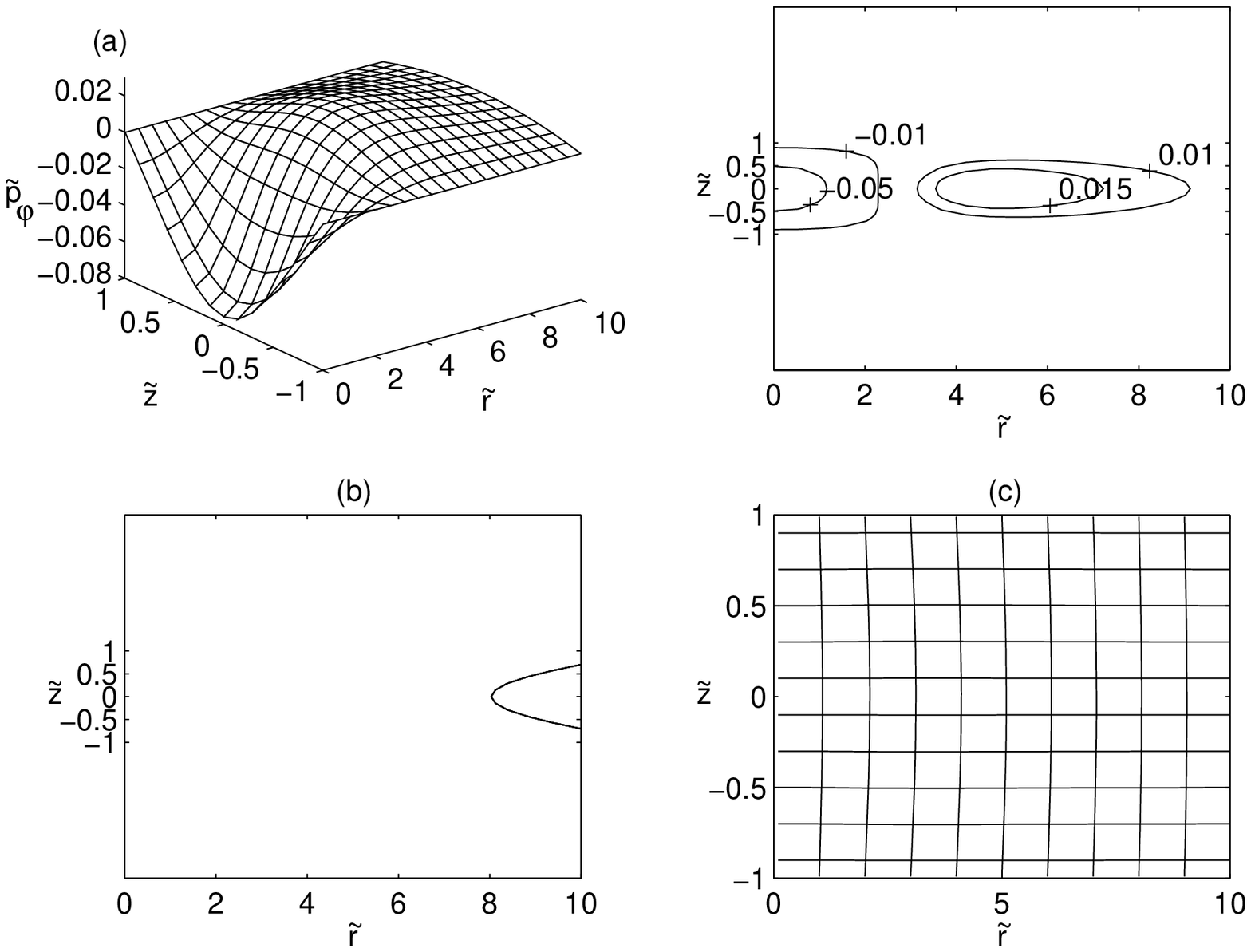}
\caption{(a) The azimuthal stresses, (b) the level curve of $|p_{\varphi}/\epsilon| =1$ and (c) the 
lines of flow calculated from Eq.\ (\ref{eq_level_z})-(\ref{eq_level_r}). 
Parameters: $\tilde{m}=1$, $\tilde{b}=5$ and $\tilde{c}=0$.} \label{fig_14}
\end{figure}

Although the expressions are exact, we do not state them explicitly, since they have
dozens of terms. The analysis must also be done graphically. We find that $p_{+}$ is a 
pressure along a direction mostly vertical and $p_{-}$ is a tension along a direction 
mostly radial. In Fig.\ \ref{fig_12}-\ref{fig_14} we graph the surfaces and level curves of the energy
density, the effective Newtonian density, the pressure mostly vertical, the tension mostly radial and  the azimuthal stresses with parameters
$\tilde{m}=1$, $\tilde{b}=5$, $n=1$ and $\tilde{c}=0$. The variables and parameters 
were rescaled as in the previous sections. While densities and the radial tension decrease monotonously 
with $\tilde{r}$ and $\tilde{z} \rightarrow \pm 1$, the upward pressure first increases with radius and then
decreases monotonously. The azimuthal stresses are negative near the disk's center, then increase 
to a positive maximum at the $\tilde{z}=0$ plane and then decrease monotonously. In some sense
this new class of disks has similar characteristics of those obtained in Weyl coordinates, but here the almost
radial tensions are different from the almost vertical pressures. Unfortunately the dominant energy condition 
does not hold with respect to the azimuthal stresses. Fig.\ \ref{fig_14}(b) shows the level curve
of $|p_{\varphi}/\epsilon| =1$. For $\tilde{r} \gtrapprox 8$ the dominant energy condition is not satisfied.
This also seems to happen for larger values of $\tilde{b}$. In Fig.\ \ref{fig_14}(c) we plot the
lines of flow calculated by numerically solving the differential equations
\begin{subequations}
\begin{align}
\frac{dV^z}{dV^r} &=\frac{V^z}{V^r} \mbox{,} \label{eq_level_z}\\
\frac{dX^z}{dX^r} &=\frac{X^z}{X^r} \mbox{.} \label{eq_level_r}
\end{align}
\end{subequations}
The vertical and horizontal lines in Fig.\ \ref{fig_14}(c) are, respectively, the lines of flow
associated to Eq.\ (\ref{eq_level_z}) and Eq.\ (\ref{eq_level_r}). We note that the lines are very
parallel to the coordinate axes. This happens because the component $T^z_r$ is about two orders 
of magnitude smaller than the other components of the energy-momentum tensor, thus 
the eigenvalues $p_{\pm}$ are almost equal to the corresponding diagonal elements. We call then the
eigenvalue $p_{+}$ with eigenvector $V^a$ the pressure
mostly vertical and the eigenvalue $p_{-}$ with eigenvector $X^a$ the tension mostly radial.

\section{Discussion} \label{sec_discuss}
The ``displace, cut, fill and reflect'' method and a class of functions used to ``fill'' was used to
construct new classes of thick general relativistic disks that generalize the models
studied in a previous work \cite{Gonzalez3}. The Newtonian Kuzmin-Toomre disks were 
used to put constraints on the parameter in the class of ``fill'' functions so that the disks were
physically acceptable. Then the Schwarzschild solution was used to construct disks 
in isotropic cylindrical coordinates, Weyl coordinates and canonical Schwarzschild coordinates. 
In isotropic coordinates the disks have equal radial and azimuthal pressures (isotropic fluid) 
but different vertical pressures. Disks in Weyl coordinates present radial tensions that are 
equal in modulus as vertical pressures, azimuthal tensions near the disk's center and 
azimuthal pressures for larger radii. All disks are in agreement with the strong, weak and 
dominant energy conditions.

In canonical Schwarzschild coordinates the ``displace, cut, fill and reflect'' method cannot 
be applied directly and an additional odd function must be added to generate exact disks.
We find that the disks have tensions along a direction mostly radial that are different from 
the pressures along a direction mostly vertical, and the azimuthal stresses have similar 
behavior as in disks in Weyl coordinates. These disks are in agreement with the strong and 
weak energy conditions, but the azimuthal stresses do not satisfy the dominant energy condition.

\bigskip
\centerline{{\large \textbf{Acknowledgments}}}

\bigskip
D.\ V.\ thanks CAPES for financial support. P.\ S.\ L.\ thanks CNPq and FAPESP 
for financial support.


\begin{thebibliography}{99}
\bibitem{Bonnor1} W. A. Bonnor and A. Sackfield, \textit{Comm. Math. Phys.} \textbf{8}, 338 (1968).
\bibitem{Morgan1} T. Morgan and L. Morgan, \textit{Phys. Rev.} \textbf{183}, 1097 (1969).
\bibitem{Morgan2} L. Morgan and T. Morgan, \textit{Phys. Rev. D}\textbf{2}, 2756 (1970).
\bibitem{Lynden-Bell} D. Lynden-Bell and S. Pineault, \textit{Mon. Not. R. Astron. Soc.} \textbf{185}, 679 (1978).
\bibitem{Letelier1} P. S. Letelier and S. R. Oliveira, \textit{J. Math. Phys.} \textbf{28}, 165 (1987).
\bibitem{Lemos1} J. P. S. Lemos, \textit{Class. Quantum Grav.} \textit{6}, 1219 (1989).
\bibitem{Lemos2} J. P. S. Lemos and P. S. Letelier, \textit{Class. Quantum Grav.} \textbf{10}, L75 (1993).
\bibitem{Bicak1} J. Bi\v{c}\'{a}k, D. Lynden-Bell and J. Katz, \textit{Phys. Rev. D}\textbf{47}, 4334 (1993).
\bibitem{Bicak2} J. Bi\v{c}\'{a}k, D. Lynden-Bell and C. Pichon, \textit{Mon. Not. R. Astron. Soc.} \textbf{265}, 126 (1993).
\bibitem{Lemos3} J. P. S. Lemos and P. S. Letelier, \textit{Phys. Rev. D}\textbf{49}, 5135 (1994).
\bibitem{Lemos4} J. P. S. Lemos and P. S. Letelier, \textit{Int. J. Mod. Phys. D}\textbf{5}, 53 (1996). 
\bibitem{Espitia} G. Gonz\'{a}lez and O. A. Espitia, \textit{Phys. Rev. D} \textbf{68}, 104028 (2003).
\bibitem{Garcia} G. Garc\'{\i}a and G. Gonz\'{a}lez, \textit{Phys. Rev. D} \textbf{69}, 124002 (2004).
\bibitem{Bicak3} J. Bi\v{c}\'{a}k and T. Ledvinka, \textit{Phys. Rev. Lett.} \textbf{71}, 1669 (1993).
\bibitem{Gonzalez1} G. Gonz\'{a}lez and P. S. Letelier, \textit{Phys. Rev. D}\textbf{62}, 064025 (2000).
\bibitem{Gonzalez2} G. Gonz\'{a}lez and P. S. Letelier, \textit{Class. Quantum Grav.} \textbf{16}, 479 (1999).
\bibitem{Letelier2} P. S. Letelier, \textit{Phys. Rev. D}\textbf{60}, 104042 (1999).
\bibitem{Katz1} J. Katz, J. Bi\v{c}\'{a}k and D. Lynden-Bell, \textit{Class. Quantum Grav.} \textbf{16}, 4023 (1999).
\bibitem{Vogt1} D. Vogt and P. S. Letelier, \textit{Phys. Rev. D}\textbf{68}, 08410 (2003).
\bibitem{Vogt2} D. Vogt and P. S. Letelier, \textit{Phys. Rev. D}\textbf{70}, 064003 (2004).
\bibitem{Karas} V. Karas, J. M. Hur\'{e} and O. Semer\'{a}k, \textit{Class. Quantum Grav.} \textbf{21}, R1 (2004).
\bibitem{Klein1} C. Klein, \textit{Class. Quantum Grav.} \textbf{14}, 2267 (1997).
\bibitem{Neugebauer} G. Neugebauer and R. Meinel, \textit{Phys. Rev. Lett.} \textbf{75}, 3046 (1995).
\bibitem{Klein2} C. Klein and O. Richter, \textit{Phys. Rev. Lett.} \textbf{83}, 2884 (1999).
\bibitem{Klein3} C. Klein, \textit{Phys. Rev. D}\textbf{63}, 064033 (2001).
\bibitem{Frauendiener} J. Frauendiener and C. Klein, \textit{Phys. Rev. D}\textbf{63}, 084025 (2001).
\bibitem{Klein4} C. Klein, \textit{Phys. Rev. D}\textbf{65}, 084029 (2002).
\bibitem{Klein5} C. Klein, \textit{Phys. Rev. D}\textbf{68}, 027501 (2003).
\bibitem{Klein6} C. Klein, \textit{Ann. Phys.} \textbf{12} (10), 599 (2003).
\bibitem{Gonzalez3} G. Gonz\'{a}lez and P. S. Letelier, \textit{Phys. Rev. D}\textbf{69}, 044013 (2004).
\bibitem{Kuzmin} G. G. Kuzmin, \textit{Astron. Zh.} \textbf{33}, 27 (1956).
\bibitem{errata1} These expressions are the generalizations of Eq.\ (30a)-(30e) of Ref.\ \cite{Gonzalez3}. 
There is a misprint in Eq.\ (30a) and (30b), where the numerical factor multiplying the equations should be 32.
\bibitem{Weyl1} H. Weyl, \textit{Ann. Phys.} \textbf{54}, 117 (1917).
\bibitem{Weyl2} H. Weyl, \textit{Ann. Phys.} \textbf{59}, 185 (1919).
\bibitem{errata2} Note that this inequality is different from condition (24) deduced in Ref.\ \cite{Gonzalez3}, because
Eq.\ (22b) in that paper has an error in the term that multiplies $\tilde{\rho}$: the product $\tilde{R}_1\tilde{R}_2$ 
in the denominator should not be squared (see Eq.\ (\ref{eq_eps_weyl})). The correct inequality is 
$0<\tilde{m}<(\tilde{b}^2-\tilde{k}^2)/\tilde{b}$.
\end{thebibliography}
\end{document}